\documentclass[aps,prl,twocolumn,showpacs,superscriptaddress,10pt]{revtex4-1}%
\usepackage{amsmath}
\usepackage{amsfonts}
\usepackage{amssymb}
\usepackage[caption=false,justification=raggedright,singlelinecheck=false]{subfig}
\usepackage{graphicx}
\usepackage[export]{adjustbox}
\usepackage{enumitem}
\usepackage{color}
\usepackage{bbold}
\usepackage{cancel}
\usepackage{hyperref}
\usepackage{bm}

\allowdisplaybreaks[4]

\newcommand{\mb}[1]{\mathbf{#1}}
\newcommand{\bq}{\mb{q}}
\newcommand{\mc}[1]{\mathcal{#1}}

\newcommand{\mean}[1]{\langle #1 \rangle}

\makeatletter
\DeclareRobustCommand{\rvec}[1]{%
  \mathpalette\do@rvec{#1}%
}
\newcommand{\do@rvec}[2]{%
  \fix@rvec{#1}{+}%
  \reflectbox{$\m@th#1\vec{\reflectbox{$\fix@rvec{#1}{-}\m@th#1#2\fix@rvec{#1}{+}$}}$}%
  \fix@rvec{#1}{-}%
}
\newcommand{\fix@rvec}[2]{%
  \ifx#1\displaystyle
    \mkern#23mu
  \else
    \ifx#1\textstyle
      \mkern#23mu
    \else
      \ifx#1\scriptstyle
        \mkern#22mu
      \else
        \mkern#22mu
      \fi
    \fi
  \fi
}
\makeatother

\makeatletter
\newcommand{\lowersim}[2]{%
  \sbox\z@{$#1<$}%
  \raisebox{-\dimexpr\height-\ht\z@}{$\m@th#1#2$}%
}
\makeatother

\begin{document}
\title{Topological Field Theory far from Equilibrium}
\author{F. Tonielli}
\affiliation{Institut f\"{u}r Theoretische Physik, Universit\"{a}t zu K\"{o}ln, D-50937 Cologne, Germany}
\author{J. C. Budich}
\affiliation{Institute of Theoretical Physics, Technische Universit\"{a}t Dresden, 01062 Dresden, Germany}
\author{A. Altland}
\affiliation{Institut f\"{u}r Theoretische Physik, Universit\"{a}t zu K\"{o}ln, D-50937 Cologne, Germany}
\author{S. Diehl}
\affiliation{Institut f\"{u}r Theoretische Physik, Universit\"{a}t zu K\"{o}ln, D-50937 Cologne, Germany}

\begin{abstract}
The observable properties of topological quantum matter are often described by topological field theories. We here demonstrate that this principle extends beyond thermal equilibrium. To this end, we construct a model of two-dimensional driven open dynamics with a Chern insulator steady state. Within a Keldysh field theory approach, we show that under mild assumptions -- particle number conservation and purity of the stationary state -- an abelian Chern-Simons theory describes its response to external perturbations. As a corollary, we predict chiral edge modes stabilized by a dissipative bulk.

\end{abstract}

\date{\today}

\maketitle
\emph{Introduction} -- The topological properties of many-body systems in zero
temperature equilibrium states are encoded in twists of their ground state wave
function \cite{Thouless1982, Thouless83, KaneMele05, HasanKane2010, QiZhang2011,
Asboth2016}. Recently, there has been increasing interest in exploring how such
structures generalize beyond  equilibrium. New concepts developed along these lines
include  Floquet topological phases \cite{Lindner2011,Rudner2013,Rudner2019},
dissipative engineering of topological states \cite{Diehl2011,Bardyn2013,Budich2015},
and topological non-Hermitian systems \cite{Weimann2017, Zhou2018, Gong2018,
Kunst2018, Kawabata2019}. These developments are motivated in part by breakthroughs
realizing out of equilibrium topological matter in  experimental platforms, such as
ultracold atoms \cite{Goldman2016}, photonic settings \cite{Lu2014,Ozawa2019}, and
exciton-polariton systems \cite{St-Jean2017,Klembt2018}. This multitude of emerging
concepts and application fields raises the questions for universal organizing
principles in the topology of matter.

In equilibrium, one such overarching framework is the topological field theory
\cite{Zhang1989, Lopez1991, LevinWen08, Qi2008} approach. Based on the interplay of
topology and gauge structures, such effective theories provide a versatile bridge
between microphysics and observable system properties~\cite{Redlich1984,Zhang1989,
Lopez1991,Ryu2012}. Where these gauge principles exist, they show a high level of
robustness, including in the presence of interactions~\cite{Ryu2012} or translational symmetry
breaking~\cite{Bagrets16}. On the same basis, they describe the connection between
bulk and boundaries, and the formation of  edge modes~\cite{Wen1995,Tong2016}.
Representative for numerous other implementations~\cite{Zhang1989, Lopez1991, LevinWen08, Qi2008,Wen2008}, the perhaps simplest example in
this category is the Chern-Simons (CS) theory describing the electromagnetic response of the (anomalous) quantum Hall
insulator~\cite{Ryu2012} by extension of an earlier construction in
$(2+1)$-dimensional quantum electrodynamics~\cite{Redlich1984}.

In this Letter we address the question whether the topological gauge response
 approach is tied to thermal equilibrium. To this end, we consider an extreme
 opposite of the Hamiltonian  quantum Hall paradigm: topology  defined by dissipative
 state engineering and absence of Hamiltonian dynamics 
 \cite{Diehl2011,Bardyn2013,Budich2015, goldstein2018, shavit2019}. 
 We start out from a quantum master equation
 of Lindblad form stabilizing a stationary point (`dark state') that is identical to
 the ground state of an anomalous quantum Hall insulator. In this way the stationary
 state and static correlation functions  coincide with those of the Hamiltonian
 ground state scenario. Yet, the dynamics steering the system into that state is
 fundamentally different: it violates  equilibrium principles such as detailed
 balance, and is dissipative instead of unitary. We will show that, despite of these differences, Chern-Simons  theory emerges as the effective response theory (cf. Eq.~\eqref{eq:CSFinal} below). In this way,
 our findings extend the scope of topological field theory to systems driven far out
 of equilibrium. Specifically, they demonstrate that quantum mechanical unitarity is not essential to the stabilization of a topological response theory.

\emph{Microscopic Lindblad model} --  
We consider the dynamics of a Markovian quantum master equation in Lindblad form \cite{Lindblad1976, breuer2002} and in the spatial continuum, 
\begin{equation}
\label{eq:KeyQME}
\hspace{-0.2cm}\frac{d}{dt}\hat{\rho}=\hspace{-0.1cm} \int_\mathbf{x}\hspace{-0.1cm}\big(\hspace{-0.05cm}-\hspace{-0.05cm}i\big[\hat{\mathcal{H}}_0,\hat{\rho}\big] + \sum_{\alpha}\gamma_{\alpha}\big[2\hat{L}_{\alpha}^{\,}\hat{\rho}\hat{L}_{\alpha}^{\dagger} - \{\hat{L}_{\alpha}^{\dagger}\hat{L}_{\alpha}^{\,},\hat{\rho}\} \big]\big),
\end{equation}
where the Lindblad operators, $\hat{L}_{\alpha}$,
generate  driven--dissipative dynamics, and $\hat{\mathcal{H}}_0$ represents the optional presence of coherent Hamiltonian dynamics. We now construct the non-equilibrium analog of a gapped ground state, by
requiring the existence of a  stationary state, $\hat \rho_D =|D\rangle\langle
D |$, $\frac{d}{dt} \hat \rho_D=0$, satisfying
\begin{align}
\label{eq:KeyDarkstate}
&\big[\hat{H}_0,\left|D\right\rangle \langle D|\,\big]=0,& &\hat{L}_{\alpha}\left|D\right\rangle = 0,
\end{align}
where $\hat{H}_0 =\int_{\mb{x}}\hat{\mathcal{H}}_0$. We require the state $|D\rangle$
and the dynamics stabilizing it to satisfy a number of defining conditions:
$|D\rangle$ should \emph{(i)} carry topological charge, \emph{(ii)} be unique such
that $\hat \rho_D$ is a pure state, and \emph{(iii)} stable in that local
perturbations to the steady state relax at a finite minimal rate, defining the
`dissipative gap' of the system. We also require \emph{(iv)} particle number
conservation of the dynamics generated by $\hat L_\alpha$, and \emph{(v)} spatial
locality of the same operators. Here, \emph{(ii)}--\emph{(iv)} implement conditions
otherwise required by Laughlin's gauge argument~\cite{Laughlin81}: the threading of a
quantum Hall annulus by a time varying magnetic flux  can be adiabatic only if the
bulk state is non-degenerate and has a many-body spectral gap. In this case, the
insertion of flux quanta will lead to the transfer of an integer number of charges
from one edge to the other, \emph{provided} these charges cannot be lost (e.g., to a
bath). In practical terms, particle number conservation implies that $\hat L_\alpha$
are (at least) quadratic in elementary particle operators, and $\hat L^\dagger_\alpha
\hat L_\alpha $ quartic \footnote{Einstein's summation convention on repeated indices
is assumed unless otherwise specified}: the model we are constructing is strongly
interacting by design.

The above criteria \emph{(i)}--\emph{(v)} are implemented in one go by defining the
jump operators $\hat L_\alpha$ in correspondence to a two-band topological insulator
model. To start with, we pick a
reference Hamiltonian, parameterized as $\hat H =\int_\bq \hat H_\bq$, with $\hat H_\bq\equiv
\hat\psi_{\bq}^\dag\, (\mb{d}_\mb{q}\cdot  \bm{\sigma})\,\hat\psi_{\bq}$. Here,
$\int_{\bq}\equiv \int \frac{d^{2}q}{(2\pi)^{2}}$,
$\hat{\psi}_{\bq}\equiv\begin{pmatrix}\hat{\psi}_{1,\bq}, 
\hat{\psi}_{2,\bq}\end{pmatrix}^{T}$ is a two-component vector, and $\bm{\sigma}=(\sigma_1,\sigma_2,\sigma_3)^T$ the vector of Pauli matrices. The specific choice $\mb{d}_\bq\equiv(2m q_1,2mq_2,-m^{2}+\bq^2)$ defines the continuum representation of a two-dimensional Chern insulator
\cite{Asboth2016}, where the winding of the map $\bq\mapsto \mb{d}_\bq$ defines the Chern number $\theta=-1$ (for any $m\not=0$).

The insulating configuration corresponds to half filling, i.e., equal occupation
density of particles and holes,
$\mean{\hat{\psi}_{\alpha}^{\,}\hat{\psi}^{\dagger}_{\alpha}}=\mean{\hat{\psi}_{\alpha}^{\dagger}\hat{\psi}^{\,}_{\alpha}}=n$, where
$n$ is a (formally diverging) factor of the order of the squared inverse lattice
spacing of a microscopically defined topological insulator with the above continuum limit \cite{Qi2008}. In this
configuration,  the ground state $|D\rangle$ of $\hat H$ is defined by the occupation
of all states with negative eigenvalue of modulus $d_\bq\equiv
|\mb{d}_\bq|=(\bq^2+m^2)$. Identifying this state with the dark state of the
dissipative dynamics, we now define a set  $\{\hat L_\alpha\}$ satisfying the above
conditions \emph{(i)}-\emph{(v)}: consider the four operators $\hat L^{\,}_{1,2}=\hat
\psi^\dagger_{1,2}\hat l^{\,}_1$, $\hat L^{\,}_{3,4}=
\hat \psi^{\,}_{1,2}\hat l^\dagger_2$, where the operators $\hat l_{1,2}$ diagonalize the Hamiltonian as $\hat H^{\,}_\bq\equiv \hat l^\dagger_\bq \sigma_3 \hat l^{\,}_\bq$ with
\begin{align}
  \label{eq:lVsPsi}
  \hat l_\bq=V_\bq\hat \psi_\bq,\quad V_\bq\equiv q_1 \sigma_0+i q_2 \sigma_3+i m \sigma_2.
\end{align}
 The matrices $V_\bq$ differ from the unitary transformations $\hat c_\bq \equiv U_\bq \hat \psi_\bq$ defining  the eigenbasis, $\hat H_\bq\equiv 
d_\bq \hat c^\dagger_\bq \sigma_3 \hat c_\bq$, by only a scalar factor $V_\bq
=
d_\bq^{\frac{1}{2}} U_\bq$, i.e. the definition of the ground state 
can equally be represented in the $\hat c$-- or
$\hat l$--representation. However, the advantage of working with the latter is that the matrices $V_\bq$ contain only one spatial derivative, $q_i =-i\partial_i$, so that the bilinears $\hat \psi^{\dagger}_{\alpha}$ are local in space, \eqref{eq:lVsPsi}, while $ \hat \psi^{\dagger}_{\alpha} \hat{c}^{\,}_{\beta}$ would be strongly nonlocal.  

\begin{figure}[!t]
  \centering
   \includegraphics[width=\columnwidth,scale=1,valign=c]{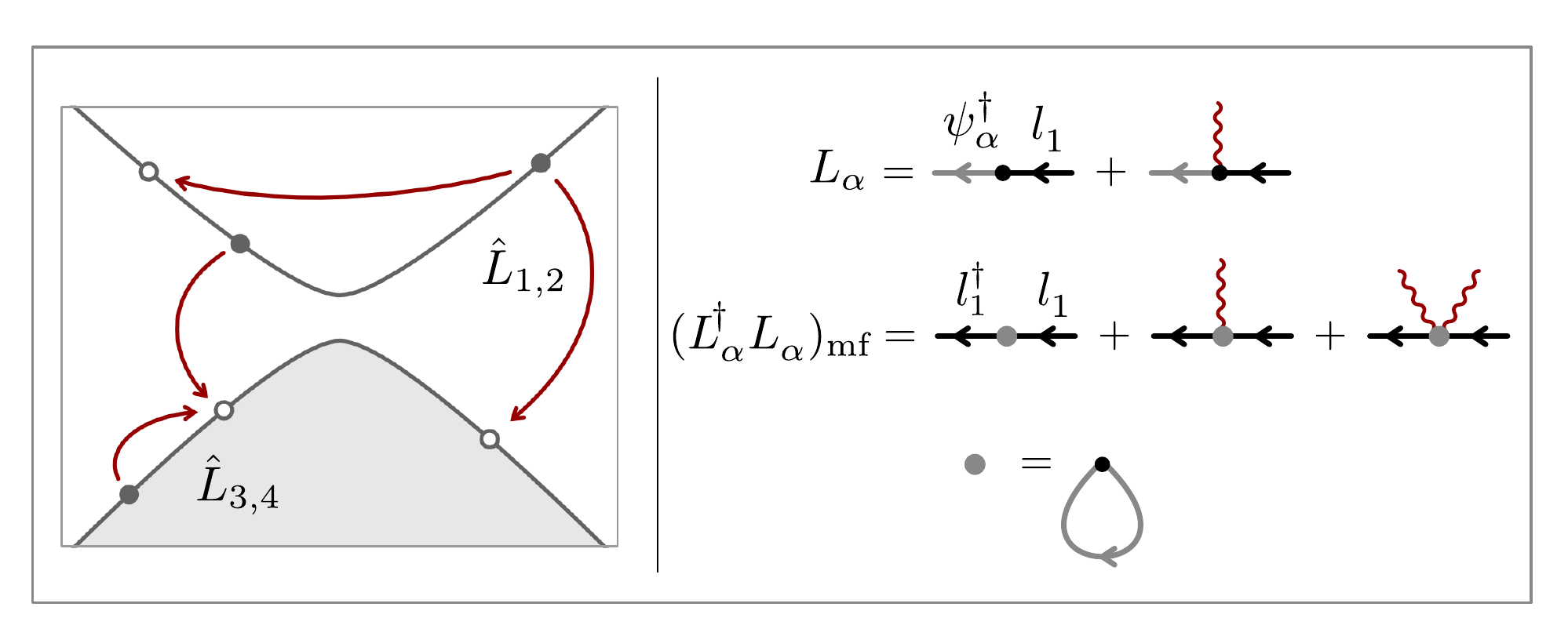}
    \caption[]{Left: Visualizing the action of the jump operators $\hat L$. The
    operators $\hat L_{1,2}$  annihilate particles in the upper band to either
    re-create them in the lower, or redistribute them in the upper band. Similarly,
    $\hat L_{3,4}$ create particles in the lower band by transfer from the upper or
    redistribution from the lower band. The stationary state of this process is a
    fully occupied lower band. Right: A mean field decoupling 
    $\langle {\psi}_{\alpha}^{\,} {\psi}_{\alpha}^{\dagger}\rangle\to n$ reduces the quartic field polynomials 
    $L^\dagger L\to l^\dagger  l $ to quadratic ones. The presence of the wavy line
    indicates that the $L$-fields carry a non-trivial gauge representation,
     upgrading them to  current operators in the presence of an external
    vector potential, cf. Eq.~\eqref{eq:LACoupling}.}
     \label{fig1}
\end{figure}

\emph{Keldysh field theory} -- Our goal is to describe the long time/distance response of
a system governed by the  dissipative dynamics \eqref{eq:KeyQME} to a
perturbation represented by an external gauge field. Rather than working with the
equation itself, we approach this task  in the language of a unit normalized  Keldysh functional integral,
\begin{eqnarray}\label{eq:Keld}
Z &=& \int \mathcal D\psi\, e^{i S[\psi ]},\\\nonumber
S[\psi] &=& \int_{t,\mb{x}}\Big[\psi^\dagger_{+}i\partial_{t}\psi_{+}-\mathcal{H}_{+}-(+\to-)\label{eq:Keldyshaction-ham}\\\nonumber
&& -i\gamma\sum_{\alpha}\Big(2L_{\alpha,+}{L}^\dagger_{\alpha,-}\label{eq:Keldyshaction-diss} -{L}^\dagger_{\alpha,+}L_{\alpha,+}-{L}^\dagger_{\alpha,-}L_{\alpha,-}\Big)\Big],
\end{eqnarray}
carrying equivalent information \cite{Kamenev2011b,Sieberer2015}. In
Eq.~\eqref{eq:Keld}, we assume identical couplings $\gamma_\alpha\equiv \gamma$ for
simplicity, $\psi=(\psi_+,\psi_-)^T$ is now a field of anticommuting Grassmann
variables,  and the shorthand notation $\mathcal{H}_\pm = \mathcal{H}(\psi_\pm),
L_\pm= L(\psi_\pm)$ is used. In Eq.~\eqref{eq:Keld}, fields  carrying a Keldysh
contour index $\pm$ assume the role of operators acting in Eq.~\eqref{eq:KeyQME} on
the density matrix from the left/right. Specifically, the quartic operators
$L^\dagger L$ define a dissipative variant of an instantaneous two-body interaction.
Via the processes illustrated in Fig.~\ref{fig1}, they drive an exponentially fast
population of the ground state of $\hat H$. This state is unique, and protected by a
 dissipative gap against the formation of long-lived excitations, e.g., of
particle--hole type (see Supplemental Material C
\setcounter{footnote}{19}\footnote{See Supplemental Material for details on the
self-consistent Born approximation, on the evaluation of the prefactor of the
Chern-Simons action, and a discussion on the existence of a many-body dissipative gap
through a paradigmatic example}). These features stabilize a mean field approach,
which is the key to progress with the strongly interacting theory
Eq.~\eqref{eq:Keld}.

An inspection of the quartic terms  shows that their leading contribution to the functional
integral, formally equivalent to a one-loop self-consistent Born approximation, comes
from replacements such as $ l_1^\dagger \psi_\alpha \psi^\dagger_\alpha
 l_1 \to  l_1^\dagger \langle \psi_\alpha
\psi^\dagger_\alpha\rangle  l_1 = n l_1^\dagger l_1$ (see Supplemental Material A
\cite{Note20} for technical details). Prior to the introduction of gauge fields, the
same decoupling applies to all terms of quartic order. In effect, it amounts to a
substitution $L^{\,}_{1,2}\to  l^{\,}_1$, $L^{\,}_{3,4}\to l^\dagger_2$, and an absorption
$\gamma\to n\gamma\equiv \bar\gamma$ of the density factor in an effective coupling
constant.

At this point, the theory has become quadratic in the fields. 
Representing the $l$-fields via Eq.~\eqref{eq:lVsPsi} through $\psi$'s and doing the Gaussian integrals describing the theory after mean field decoupling, we obtain the retarded and Keldysh Green's functions \footnote{In terms of contour fields, suppressing all arguments but time for simplicity, they are equal to {$G^{R}(t) =  \theta(t)(G^{>}(t)-G^{<}(t))$} and {$G^{K}(t) = G^{>}(t)-G^{<}(t)$}, with {$iG^{>/<}_{\alpha\beta}(t) = \mean{\psi_{\mp,\alpha}^{\,}(t)\psi_{\pm,\beta}^{\dagger}(0)}$} \cite{Kamenev2011b}\label{Keldyshdefinition}} 
($\gamma_\bq  =  \bar \gamma d_\bq$)
\begin{align}
\label{eq:M1gf}
G^{K}_{\omega,\mb{q}}\,=\,-2i\bar\gamma\frac{ {\mb{d}_\bq\cdot\bm{\sigma}}}{\omega^{2} + \gamma_\bq^{2}},\quad G^R_{\omega,\mb{q}}\,=\, \frac{\mathbb{1}}{\omega + i\gamma_\bq }\, .
\end{align}
Here, the different matrix structure of $G^{R}$ and $G^{K}$ implies the absence of a
thermodynamic fluctuation--dissipation relation \cite{Kamenev2011b, Sieberer2015}. Moreover, the information
on the topological band structure abides in $G^K$ (via the matrix  {$\mb{d}_\bq\cdot\bm{\sigma}$}), while $G^R$ knows only about the
spectral structure through  the function $\gamma_\bq$. The structure of the Green's function also shows that in mean field theory the many body dissipative damping mechanism reduces to a spectral  gap for single-particle excitations, $i \gamma_\bq\stackrel{\bq\rightarrow0}\longrightarrow i \bar \gamma m^2$. Therefore, both, single--particle and particle-hole excitations are gapped out.




\emph{Gauge theory} -- In its present form, the theory describes the relaxation of generic states $\hat
\rho$ into the Chern insulator dark state $\hat \rho_D$. We now take the next step to couple the
fermions to a gauge field and in this way probe the response to 
external perturbation. To this end, we go back to the original Keldysh
action~\eqref{eq:Keld} and notice that it possesses a $\mathrm{U}(1)\times
\mathrm{U}(1)$ symmetry under independent phase rotations $\psi_\sigma \to
e^{i\chi_\sigma}\psi_\sigma$, $\sigma=\pm$ of the fields on the two contours. On
general grounds, phase rotations with spatio-temporal variation generate a finite
action cost where $\sum_{\sigma}\int_{t,\mb{x}} \sigma \partial_\mu \chi_\sigma
J^\mu_\sigma$, ($\partial_\mu = (\partial_t, \vec \nabla)$), and $J^\mu_\sigma$,
define conserved currents of the theory \cite{Sieberer2015}.
The symmetry under phase rotations is upgraded to a local one by gauging it  {\cite{Avron2011}}. We do so
by  minimally coupling  occurrences of phase gradients in Eq.~\eqref{eq:Keld} 
 as $\partial_\mu \chi_\sigma\to \partial_\mu \chi_\sigma+
A_{\sigma,\mu}$ to the components of a vector potential, independently for both
contours. In this way, $Z\to Z[A]$ becomes a sourced functional, from which
expectation values of currents can be computed as derivatives. Of particular interest are the elements of the DC conductance tensor \cite{Mahan}, $\sigma_{ij}=\lim_{\omega,\bq\to 0} \omega^{-1}\delta^2_{ A_{q,i,\bq,\omega},A_{c,j,-\bq,-\omega}}Z[A]$ , where the Keldysh representation $A_c=(A_++A_-)/2, A_q=A_+-A_-$ is used. 

To give these expressions concrete meaning, the coupling of the gauge field to the
action needs to be made explicit. From Eq.~\eqref{eq:Keld}, we infer that the
temporal component couples to the action as $\int_{t,\mb{x}} \psi^\dagger_\sigma A_{\sigma,0}
\psi_\sigma$. The coupling to the spatial components is more interesting, and this is
where the interplay of topology and dissipation comes in: consider the jump operator
$L_1=\psi^\dagger_1 l_1$. With $l=V\psi$, phase transformations affect this
expression as $L_{1,\sigma}\to L_{1,\sigma}+ \psi_1^\dagger (\partial_{q_{i}}VV^{-1} l)_1
\,\partial_i \chi$, where  $\partial_{q_{i}} V V^{-1}$ is a matrix
local in momentum space, but  non-local in real space. Using that $V_\bq$ and the unitary
diagonalizing matrices $U_\bq=d_\bq^{\frac{1}{2}}V_{\bq}$ differ only by a scalar
factor, we find that, up to an inessential diagonal matrix, $\partial_{q^i}V
V^{-1}=-ia_i+\dots$, where $a_i \equiv
i\partial_{q_i}U U^{-1}$ is the Berry connection defining the topology of the
system \cite{Asboth2016}. In this way we conclude that 
\begin{align}
\label{eq:LACoupling}
L_{1}\to L_1 -i \psi_1^\dagger  (a_i l)_1A_i
\end{align}
describes the minimal coupling of the jump operators to both the external gauge
field $A_i$, and the `internal' gauge field $a_i$ (cf. wavy line in the top row of Fig.~\ref{fig1} left.)  With this substitution the bilinears $L_1^\dagger L_1^{\,}$ pick up $A$-dependence of up to second order.

\begin{figure}[!t]
  \centering
   \includegraphics[width=\columnwidth,scale=1,valign=c]{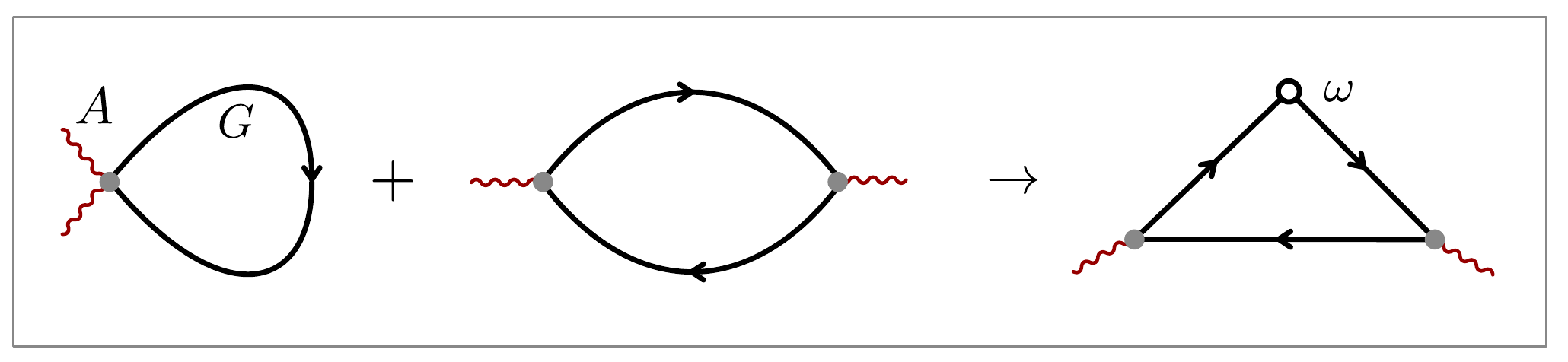}
    \caption[]{Structure of the diagrams appearing in the expansion in $A$. Both diagrams are individually ultraviolet-divergent in momentum space. The divergences cancel and the remaining contribution is identical to the triangle structure on the right, where the empty dot represents the expansion of the propagators $G$ to first order in the external frequency, $\omega$. This gives a contribution $\sim \omega A_i A_j$ which becomes one part of the Chern-Simons action.}
     \label{fig2}
\end{figure}
We next expand the action to second order in $A_i$ and apply the same mean field
decoupling as that outlined above. This procedure is equivalent to a one-loop
approximation of the action (cf. Fig.~\ref{fig2},  and Supplemental Material B
\cite{Note20} for details). It is reminiscent of the computation of an induced CS
term \cite{Redlich1985, dunne1999} in $(2+1)$-dimensional quantum electrodynamics by
loop expansion, later proven to be  {unchanged} under the inclusion of higher order gauge
invariant  {gauge}--matter interactions \cite{Coleman1985}. Under the condition
of purity of the dark state, which is met by the model
\eqref{eq:KeyDarkstate}-\eqref{eq:lVsPsi} \footnote{Violations of this condition
\cite{us} play a role similar to that of finite temperatures in Hamiltonian settings
\cite{dunne1999} and may compromise the form of the Chern-Simons theory.}, this
procedure yields  $S[A]=
\frac{\theta}{4\pi} \int_{t,\mb{x}} \epsilon_{ij} A_i \partial_t A_j$, where the
prefactor is given by the Chern number of the filled band, $\theta=-\frac{\pi}{2}\int_{\bq} \text{Tr}[\sigma^{z}(\partial_{q_{1}} a_2-\partial_{q_{2}} a_1)]$. The above linear response relation shows that this parameter defines the quantized transverse conductance  as
$\sigma_{12}=\theta/2\pi$.  For the particular
two-band model defined above, the identification of $a$ through the
transformation matrices~\eqref{eq:lVsPsi} leads to the winding number representation 
 {$\theta=\frac{1}{4\pi}\int d^{2}q\,\frac{1}{d^3}\,\mb{d}\cdot (\partial_{q_{1}} \mb{d}\times\partial_{q_{2}} \mb{d})$}, 
which evaluates to the Chern number $\theta=-1$. 

\emph{Topological gauge theory} -- One may now complete the derivation of the topological action by double expansion in
the temporal and spatial components $\sim A_0 A_i$. However, there is no need to do
so explicitly, because the structure of the ensuing \emph{dissipative Chern-Simons
action} is entirely fixed by symmetry and topological principles. To see how, we note that the most general form of a Chern-Simons  action of two gauge fields $(A_c,A_q)$ reads $S_{\text{CS}}[A]=\int_{t,\mb{x}}  {M^{IJ}}A_I d A_J$,
where we switched to a compact differential form notation
$\epsilon^{\mu\nu\rho}\,A_{\mu}\partial_{\nu}A_{\rho}\rightarrow A d A$, and $M$ is
a $2\times2$ matrix. Probability conservation (equivalent to the absence of purely
$c$ contributions to the action of bosonic Keldysh theory \cite{Kamenev2011b}) requires $ {M^{cc}}=0$.
Similarly, the preserved Hermiticity of a density matrix under evolution by the
Keldysh functional requires \footnote{Despite the lack of thermal symmetry \cite{Sieberer2015}, the action \eqref{eq:Keld} is invariant under another discrete transformation, namely {$\psi_{\pm}^{\,} \to i\psi^{\dagger}_{\mp}$}, {$\psi^{\dagger}_{\pm} \to i\psi^{\,}_{\mp}$}, {$S \to -S^*$}, where the latter symbol denotes complex conjugation of the coefficients of the action. This is the field theoretic counterpart of the action of Hermitian conjugation on the Liouvillian. Under this transformation, {$S[A_{\pm}] \to -S^*[A_{\mp}]$}, from which the conditions on {$M^{IJ}$} follow.}  $S^*[A_{c},A_{q}]=-S[A_{c},-A_{q}]$, from which
$ {M^{cq}=(M^{cq})^{*}\equiv \tilde M}$ and $ {M^{qq}=-(M^{qq})^{*}}$. The condition that
topological actions enter a theory as purely imaginary phases \cite{Altland2010Condensed} in combination with
the reality of the vector potential $A$  enforces $M_{qq}=0$. We thus conclude that
the most general form of the action consistent with symmetries reads as $S_{\text{CS}}[A]=\tilde M\int_{t,\mb{x}} (A_c d A_q+A_q dA_c)$. The
\emph{quantization} of $\tilde M$ likewise follows from trace preservation, i.e., probability conservation, but in somewhat different
ways \footnote{While the quantization of coupling constants in non-abelian CS theory
is a straightforward consequence of gauge invariance, the situation in abelian
theories is somewhat more tricky \cite{Tong2016} and depends on the topology of the integration
manifold.}: the latter requires that at times $t=\pm \infty$ the Keldysh time contour
be closed, which means that time is effectively
defined on a circle. Now consider the effect of a gauge transformation, $\psi \to
e^{i\chi} \psi$, where $\chi$ is spatially constant and changes uniformly in time to
accumulate an integer winding number $2\pi W$ upon completion of the full time
revolution. In Keldysh theory the condition that such \emph{large} gauge
transformations be inconsequential enforces the quantization of observables \cite{Altland2009}, and in
the present context that of the CS coupling constant. Folding time onto the standard
forward and backward contour, the gauge transformation $A\to A + d\chi$ leaves $A_c$ invariant, while $A_{q,0} \to A_{q,0} + \frac{2\pi
W}{\Delta T}$, where $\Delta T$ is the diverging extent of the Keldysh time
interval. Substitution into the action shows that the latter changes by $\frac{2M}{\Delta T}\int_{\Delta T} dt  \Phi$, where $\Phi\equiv  \int_\textbf{x} (\partial_i A_j
- \partial_j A_i)$ is the out of plane magnetic flux through the system. Requiring
(Dirac monopole) quantization of the latter, $\Phi = 2\pi n$, on a boundary--less
spatial domain \cite{Tong2016}, we find that the gauge transformation changes the action by an
inconsequential multiple of $2\pi$, provided $M=\theta/4\pi$, with integer $\theta$. In agreement with this general argument, the calculation valid for the present model determines $\theta$ as a Chern number, thus respecting the condition. 

Summarizing, the above construction identifies
\begin{align}
  \label{eq:CSFinal}
   S_{\text{CS}}[A]&=\frac{\theta}{4\pi}\int_{t,\mb{x}} \hspace{-0.05cm} {\epsilon_{\mu\nu\rho}}\,(A_{c,\mu}\partial_{\nu}A_{q,\rho}+A_{q,\mu}\partial_{\nu}A_{c,\rho}),\cr
   \theta &= -\frac{\pi}{2}\int_{\bq} \text{Tr}[\sigma^{z}F],\quad F=\partial_{q_{1}} a_2-\partial_{q_{2}} a_1,
\end{align}
as the final form of the topological field theory describing the describing the long-time/distance response of a purely dissipative system with a pure Chern insulator steady state characterized by the Chern number $\theta$.

\emph{Boundary theory} -- In the presence of a system boundary, the Chern-Simons
theory~\eqref{eq:CSFinal} lacks gauge invariance \cite{Tong2016}. In principle, one
may attempt to identify a supplementary boundary theory compensating for this
non-invariance by microscopic construction. However, we here adopt the more
economical strategy \cite{Wen1995} to reason that  gauge invariance is restored
if the boundaries harbor a postulated gapless chiral boson mode. The minimal action
of this mode reads \cite{Wen1995} $S_0[\phi]\equiv - \frac{\theta}{4\pi}\int_{x,t}
(\partial_t \phi_c
\partial_x \phi_q+ (q\leftrightarrow c))$, where $x$ is the boundary coordinate. In
the presence of an external vector potential, $A_I$, this action picks up an
additional contribution~\cite{Stone1991} $\delta
S[\phi,A]\equiv-\frac{\theta}{4\pi}\int_{x,t}(2\partial_x \phi_c
A_{q,0}+2A_{c,0}A_{q,0}+ (q\leftrightarrow c))$. Gauge transformations, $A_I \to A_I
- \partial \chi_I, \phi_I\rightarrow \phi_I + \chi_I$ then affect the full action
$S[\phi,A]= S_0[\phi]+\delta S[\phi,A]$, in such a way that the full action
$S[\phi,A]+S_{\mathrm{CS}}[A]$ is gauge invariant. In the absence of the external
field, the boundary particle density is given by
$n(x,t)=\delta_{A_{q,0}(x,t)}\big|_{A=0} S[\phi,A]=\frac{\theta}{4\pi}\partial_x
\phi(x,t)$. The above action is minimal in that variation of $S_0$ leads to
stationarity of the boundary density $\partial_t n(x,t)=0$. To add dynamics to $n$,
elements outside pure CS theory need to be invoked. Specifically, the so far 
neglected Hermitian Hamiltonian $\hat H_0$ will generate unitary time evolution
through the operator equation $i \partial_t \hat n = -[\hat H_0,\hat n]$. For
example, if $\hat H_0$ describes a topological insulator, this leads to chiral
boundary evolution, $\partial_t \hat n = v \hat n$, with a non-universal velocity $v$. In
the boundary theory, this is accounted for by generalization $S_0[\phi]\rightarrow -
\frac{\theta}{4\pi}\int_{t,x} ((\partial_t \phi_c-v\partial_x \phi_c)
\partial_x \phi_q+ (q\leftrightarrow c))$. However, irrespective of the detailed realization of the dynamics,  the CS action generated by the dissipative bulk, requires the presence of a gapless boundary mode.

\emph{Conclusions and Outlook} -- We have considered  quantum matter  defined  via a
dissipative driving protocol with a topologically twisted dark state. This setting is
an antipode to that in  topological insulators, where mathematically identical twists
are inscribed into the  ground state of a non-interacting Hamiltonian. Our main
result is that  Chern-Simons theory emerges in either case, underpinning  the universality
of topological gauge theory. In the driven framework, its stabilization rests on
three  prerequisites: particle number conservation (formally equivalent to a double
$\mathrm{U}(1)\times
\mathrm{U}(1)$ symmetry separately for the forward and backward time evolution), purity of the dark state, and presence of a dissipative gap. Crucially, however,  quantum mechanical unitarity is nowhere required to stabilize the topological response theory.

Finally, one may look at the situation from the perspective of general geometric
response theory for Lindbladian dynamics whose formal framework has been developed in
the seminal work~\cite{Avron2011, Avron2012, Avron2012b, Fraas2016}. The present
study demonstrates how such structures materialize in concrete settings where
nonlinear  {fermion} dynamics stabilizes a system, and a minimal coupling scheme probes it.
Given that response theories define an `interface' between the micro-- and the
macrophysics of a system, this construction  may provide  useful guiding principles
to the   description  of topologically ordered quantum matter beyond the Hamiltonian
ground state setting. Specifically, one may consider the extension to  {other classes of non-Hermitian 
systems currently under active research \cite{Weimann2017, Zhou2018, Gong2018,
Kunst2018, Kawabata2019}}, and strongly entangled out of equilibrium systems with fractional excitations
\cite{B02Schuster2019}.

\emph{Acknowledgements} -- We thank M. Fleischhauer, M. Goldstein, H. Hansson, S. Moroz and M. Rudner for insightful discussions. We acknowledge support from the Deutsche Forschungsgemeinschaft (DFG, German Research Foundation) under Germany's Excellence Strategy Cluster of Excellence Matter and Light for Quantum Computing (ML4Q) EXC 2004/1 390534769, and by the DFG Collaborative Research Center (CRC) 183 Project No. 277101999 - project B02. S.D. and F.T.  acknowledge support by the European Research Council (ERC) under the Horizon 2020 research and innovation program, Grant Agreement No. 647434 (DOQS). J.C.B. acknowledges financial support from the DFG through SFB 1143 (project-id 247310070) and the Wuerzburg-Dresden Cluster of Excellence ct.qmat (EXC 2147, project-id 39085490).

\section{Supplemental material}

\subsection{A.\qquad Self-Consistent Born Approximation}
\label{app:diagrammatics}
%

The mean field approximation applied to the dissipative model is analogous to the one-loop self-consistent Born approximation developed for Hamiltonian systems \cite{Altland2010Condensed}: each quartic vertex is replaced by sums of bilinears constructed by selecting one couple of fields from the vertex and contracting the other two. For example, for the $++$ vertices involving $l_{1}$, it consists in the replacement
\begin{align}
\label{eq:HFvertex}
l^{\dagger}_{1,+}\psi^{\,}_{\alpha,+}\psi_{\alpha,+}^{\dagger}l_{1,+}^{\,}\to &-\mean{l^{\dagger}_{1,+}\psi^{\,}_{\alpha,+}}\psi^{\dagger}_{\alpha,+}l^{\,}_{1,+}\\
&+\mean{\psi^{\dagger}_{\alpha,+}\psi^{\,}_{\alpha,+}}l^{\dagger}_{1,+}l^{\,}_{1,+}+\,_{\cdots}\,,\notag
\end{align}
where Einstein's summation convention is assumed as in the main text, unless otherwise specified. A diagrammatic representation of this procedure is depicted in Fig.~\ref{fig1} in the main text.

Since the vertex is local, all contractions involve fields with the same time (and space) arguments, and the corresponding Green's functions are singular and need to be regularized by means of a point splitting of the fields. This can be done e.g.~as in \cite{Sieberer2015}, by introducing an infinitesimal correction to perfect Markovianity. 

On the $\pm$ basis, the regularization is needed only for vertices $++$ and $--$, because Green's functions with crossed indices, $\mean{\psi^{\,}_{\pm,\alpha}\psi^{\dagger}_{\mp,\beta}} = iG^{</>}_{\alpha\beta}$ \cite{Kamenev2011b}, are well-defined also for equal time arguments. For any model described by a Lindbladian, the point-splitting reads $L^{\dagger}_{\pm}(t)L^{\,}_{\pm}(t)\to L^{\dagger}_{\pm}(t\pm\varepsilon)L^{\,}_{\pm}(t)$; the same scheme is applied within each Lindblad operator, e.g., $L_{\pm} = \psi^{\dagger}_{\pm}l^{\,}_{\pm}\to \psi^{\dagger}_{\pm}(t\pm\varepsilon) l^{\,}_{\pm}(t)$. We denote the split by the superscript ${\,}^{(\varepsilon)}$ for brevity. Time/anti-time ordered Green's functions are now well-defined; for example, in Eq.~\eqref{eq:HFvertex},
\begin{align}
\label{eq:splittingGF}
\mean{\psi_{+,\alpha}^{(\varepsilon)}\psi^{\dagger}_{+, \alpha}} = iG^{\mathbb{T}}_{\alpha\alpha}(t=+\varepsilon)=\mean{\hat{\psi}_{\alpha}^{\,}\hat{\psi}^{\dagger}_{\alpha}}.
\end{align}
This is a static expectation value of operators, the order of which is \emph{fixed}. Similar equalities hold for Green's functions involving all other field and branch index combinations. In particular, it can be shown that all Green's functions of the same fields but different branch indices \emph{correspond to the same operatorial expression}, thus reducing the number of independent mean field parameters.

Contractions in Eq.~\eqref{eq:HFvertex} can now be determined. One is already computed in Eq.~\eqref{eq:splittingGF} and is fixed by the half filling condition (see main text), $\mean{\psi_{+,\alpha}^{(\varepsilon)}\,\psi^{\dagger}_{+,\alpha}} =n$.
The others must be found self-consistently, and read:
\begin{align}
\label{eq:mfparameterkC}
\mean{l^{\dagger\,(\varepsilon)}_{+,1}\,l^{\,}_{+,1}}=&\, \mean{\hat{l}^{\dagger}_{1}\hat{l}^{\,}_{1}},\\
\mean{\psi^{\dagger\,(\varepsilon)}_{+,\alpha}l^{\,}_{+,1}} = &\,\mean{\hat{\psi}^{\dagger}_{\alpha}\hat{l}^{\,}_{1}} ,\quad
\mean{l^{\dagger\,(\varepsilon)}_{+,1}\, \psi^{\,}_{+,\alpha}} = \mean{\hat{l}^{\dagger}_{1}\hat{\psi}^{\,}_{\alpha}}. \notag
\end{align}

The right hand side of Eqs.~\eqref{eq:mfparameterkC} vanishes if computed on the dark state. This is a feature shared by the analogous contractions coming from all other vertices. Setting them to zero, thus leaving $\mean{\psi^{\,}_{\alpha}\psi^{\dagger}_{\alpha}}=n$ as the only nonvanishing contraction, corresponds to the replacement $L\to l,\ \gamma\to\gamma n = \bar{\gamma}$ in the original action, and the resulting model does indeed share the same dark state as the strongly interacting one, fulfilling the self-consistent condition. A more detailed analysis \cite{us} shows that this is the only possible solution, completing the derivation.

\subsection{B.\qquad Chern-Simons level}

We substantiate here the content of Fig.~\ref{fig2} in the main text, and we also derive Eq.~\eqref{eq:CSFinal}, starting from the minimal coupling of the strongly interacting model outlined in the main text. In particular, these results show that the mean field approximation preserves gauge invariance, at least up to first order of the derivative expansion of the gauge action. We proceed on two lines: on one hand, we compute the ultraviolet divergent contributions coming from \emph{all} the vertices, to prove that the sum vanishes; on the other, we compute only the finite terms necessary to show Eq.~\eqref{eq:CSFinal}.  We focus for simplicity on the case of a purely spatial gauge field configuration, $A_{\sigma,\mu}=(0, A_{\sigma,i})$; although more involved, the case $A_{\sigma,0}\neq 0$ can be treated in complete analogy.

We recall from the main text that the model couples to the spatial components of the gauge field through shifts of Lindblad operators, expressed by Eq.~\eqref{eq:LACoupling} for $L_{1}$ and by analogous equations for $L_{2,3,4}$. The minimally coupled action has the form $S[\psi,A]=\sum_{i=0}^2 X^{(i)}$, each $X^{(i)}$ being of $i$th order in $A$ and (up to) quartic in fermionic operators. The gauge action can be obtained by integrating over fermionic degrees of freedom, with action $S[\psi]=X^{(0)}$. Denoting by $S^{(i)}[A]$ the sector of the gauge action of $i$th order in $A$, we get in cumulant expansion in second order in $A$:
\begin{subequations}
\label{eq:SA}
\begin{align}
S^{(1)}[A] &= \mean{X^{(1)}},\label{eq:SA1}\\
S^{(2)}[A] &= \mean{X^{(2)}}+\frac{i}{2}\mean{X^{(1)}X^{(1)}}_{c.}, \label{eq:SA2}
\end{align}
\end{subequations}
the subscript $c.$ in Eq.~\eqref{eq:SA2} denoting the connected part of the correlation function, $\mean{X^{(1)}X^{(1)}}_{c.}=\mean{X^{(1)}X^{(1)}}-\mean{X^{(1)}}^{2}$. We work out the full, local contribution of $\mean{X^{(1)}}$ and $\mean{X^{(2)}}$ to $S[A]$, whereas an expansion in derivatives of the fields is enough to extract the local and Chern-Simons terms from $\mean{X^{(1)}X^{(1)}}_{c.}$. More concretely, parametrizing the latter in the Keldysh representation as
\begin{align}
\label{eq:defPi}
i\mean{X^{(1)}X^{(1)}}_{c.}\equiv\int_{\omega,\bq} A^{\,}_{I,i}\Pi^{IJ}_{ij}A^{\,}_{J,j},& &I,J\in\{c,q\},
\end{align}
we determine only  $\Pi^{IJ}_{ij}(0,\mb{0})$ and $\partial_{\omega}\Pi^{cq}_{ij}(0,\mb{0})$.

Recalling from the main text the definition of the Berry connection, $a_{i}=i\partial_{q_{i}}U U^{-1}$, and suppressing the Keldysh structure for simplicity, $X^{(i)}$ are integrals of the linear combinations of the following terms:
\begin{subequations}
\label{eq:MCterms}
\begin{align}
X^{(0)} =\ &L^{\dagger}L^{\,}, \psi^{\dagger}i\partial_{t}\psi,\\
X^{(1)} =\ & l^{\dagger}_{1}\,\psi^{\,}_{\alpha}\psi^{\dagger\,}_{\alpha}\,(a^{\,}_{i\,}l)^{\,}_{1}\,A^{\,}_{i}\,,\,A^{\,}_{i}\,(a^{\,}_{i\,}l)^{\dagger}_{1}\,\psi^{\,}_{\alpha}\psi^{\dagger\,}_{\alpha}\,l^{\,}_{1}\, ,\label{eq:MCtermX1}\\
&l^{\,}_{2}\,\psi^{\dagger\,}_{\alpha}\psi^{\,}_{\alpha}\,(a^{\,}_{i\,}l)_{2}^{\dagger}\,A^{\,}_{i}\,, \,A^{\,}_{i}\,(a^{\,}_{i\,}l)^{\,}_{2}\,\psi^{\dagger\,}_{\alpha}\psi^{\,}_{\alpha}\,l^{\dagger}_{2}\,,\notag\\
X^{(2)}=\ &A^{\,}_{i}\,(a^{\,}_{i\,}l)^{\dagger}_{1}\,\psi^{\,}_{\alpha}\psi^{\dagger\,}_{\alpha} \,(a^{\,}_{j}l)^{\,}_{1}\,A^{\,}_{j}\,, \label{eq:MCtermX2}\\
&A^{\,}_{i}\,(a^{\,}_{i\,}l)_{2}^{\,}\,\psi^{\dagger\,}_{\alpha}\psi^{\,}_{\alpha} \,(a^{\,}_{j}l)_{2}^{\dagger}\,A^{\,}_{j}\,,
\notag
\end{align}
\end{subequations}
where the order of fields reflects the operatorial ordering of the corresponding terms in the Liouvillian via the point-splitting explained in Sec.~A. In Eqs.~\eqref{eq:MCtermX1} and \eqref{eq:MCtermX2}, the gauge fields $A$ have the same contour index $\sigma$ as the operator they are closest in Eqs.~\eqref{eq:MCterms} to; for example, $A_{i}^{\,}l^{\dagger}{\,}_{\cdots}\equiv A^{\,}_{\sigma,i}l^{\dagger}_{\sigma}{\,}_{\cdots}$, where $\sigma$ is a \emph{free} and fixed branch index in this case. Moreover, each curly bracket is expanded in band and momentum space as $(a_{i}l)_{\alpha,\bq} = a_{i,\alpha\beta,\mb{q}\,}l_{\beta,\mb{q}}$, with $\alpha,\beta\in\{1,2\}$ band indices.

Without any approximation, we can already infer from Eq.~\eqref{eq:MCtermX1} that $S^{(1)}[A]$ vanishes due to the dark state property, in agreement with the requirement of gauge invariance. In fact, all terms in $X^{(1)}$ either have $l_{1}^{\,}$ or $l^{\dagger}_{2}$ on the right, or their Hermitian conjugates on the left.  In both cases, they act directly on the dark state, annihilating it.

To compute the more interesting quadratic sector $S^{(2)}[A]$, that includes the topological action, we adopt the same decoupling scheme employed to obtain the Green's functions \eqref{eq:M1gf}. We make in Eqs.~\eqref{eq:MCtermX1} and \eqref{eq:MCtermX2} the replacements $\psi^{\,}_{\alpha}\psi^{\dagger\,}_{\alpha}\to \mean{\psi^{\,}_{\alpha}\psi^{\dagger\,}_{\alpha}}=n$ and $\psi^{\dagger\,}_{\alpha}\psi^{\,}_{\alpha}\to \mean{\psi^{\dagger\,}_{\alpha}\psi^{\,}_{\alpha}}=n$. 

The main building blocks of the calculations are the static and dynamic correlation functions of the eigenoperators. The former are easier to determine by adopting the point splitting procedure on the basis of contour fields, see Sec.~A. They read:
\begin{align}
\label{eq:staticGFeigenoperators}
\begin{split}
\mean{l^{\,}_{\sigma,\alpha,\mb{p}}l^{\dagger}_{\sigma',\alpha',\mb{p'}}}=d_{\mb{p}}\delta_{\alpha,1}\delta_{\alpha',1}\delta_{\mb{p},\mb{p}'},\\
\mean{l^{\dagger}_{\sigma,\alpha,\mb{p}}l^{\,}_{\sigma',\alpha',\mb{p'}}}=d_{\mb{p}}\delta_{\alpha,2}\delta_{\alpha',2}\delta_{\mb{p},\mb{p}'}.
\end{split}
\end{align}
When the correlation functions are computed at two different times, we find the Keldysh representation more practical. We denote the former by $\tilde G\sim -i \mean{l\, l^{\dagger}}$ to stress the different field combination as compared to Eq.~\eqref{eq:M1gf}. Keldysh and retarded Green's functions can be either defined as in the main text, replacing $\psi\to l$, or more compactly after the rotation \cite{Kamenev2011b} $l^{1/2} = (l^+ \pm l^{-})/\sqrt 2$ and $ l^{1/2\dagger} = (l^{+\dagger} \mp l^{+\dagger})/\sqrt 2$. In this case they can be computed as $i\tilde G^{K}_{\alpha\beta}(t) = \mean{\psi^{1}_{\alpha}\psi^{2\dagger}_{\beta}}$ and $i\tilde G^{R}_{\alpha\beta}(t) = \mean{\psi^{1}_{\alpha}\psi^{2\dagger}_{\beta}}$. From Eq.~\eqref{eq:lVsPsi} for $l_{1/2}$ and Eq.~\eqref{eq:M1gf} for $G^{K/R}$, in the frequency and momentum domains we have
\begin{align}
&\tilde{G}^{K}_{\omega,\mb{q}} = V^{\,}_{\bq}G^{K}_{\omega,\mb{q}}V^{\dagger}_{\bq}=\frac{-2i\bar{\gamma} d_{\mb{q}}^{2}\sigma^{z}}{\omega^{2}+\gamma_{\mb{q}}^{2}} \equiv  g^{K}_{\omega,\mb{q}}\sigma^{z}, \\
&\tilde{G}^{R}_{\omega,\mb{q}}= V^{\,}_{\bq}G^{R}_{\omega,\mb{q}}V^{\dagger}_{\bq}=\frac{d_{\mb{q}}\mathbb{1}}{\omega + i \gamma_{\bq}}\equiv  g^{R}_{\omega,\mb{q}}\mathbb{1}. \notag
\end{align}
We can now proceed to determine $S^{(2)}[A]$. We denote the decoupled quadratic and linear terms in the gauge field respectively by $X^{(2)}_{m.f.}$ and $X^{(1)}_{m.f.}$. The first contribution corresponds to the first diagram in Fig.~\ref{fig2}. Expressing it in terms of $A_{c}$ and $A_{q}$, we get:
\begin{align}
\label{eq:X2ev}
\mean{X^{(2)}_{m.f.}}&=i\bar{\gamma}\lambda_{ij}\int_{t,\mb{x}}\!A^{\,}_{q,i}A^{\,}_{q,j}-A^{\,}_{c,i}A^{\,}_{q,j}+A^{\,}_{q,i}A^{\,}_{c,j},\\
\label{eq:lambda}
\lambda_{ij}&\equiv\int_{\mb{p}}\big\langle(a^{\,}_{i\,}l)^{\dagger}_{1,\mb{p}}(a^{\,}_{j}l)^{\,}_{1,\mb{p}}+(a^{\,}_{i\,}l)_{2,\mb{p}}^{\,}(a^{\,}_{j}l)_{2,\mb{p}}^{\dagger}\big\rangle\,\\
&=\int_{\mb{p}}\frac{d_{\mb{p}}}{2}\,\text{Tr}\big[\big(a_{i}a_{j}\big)_{\mb{p}} -\sigma^{z}a_{i,\mb{p}}\sigma^{z}a_{j,\mb{p}}\notag\\
&\quad\quad\quad\quad\quad\  - \sigma^{z}\big(a_{i}a_{j}-a_{j}a_{i}\big)_{\mb{p}\,}\big],\notag
\end{align}
where $\text{Tr}$ denotes the trace over band indices. The static expectation values in Eq.~\eqref{eq:lambda} involve contour fields, since they are generated independently by each term of the quantum master equation. However, the respective indices are not specified because the expectation values are independent of them, as shown by Eqs.~\eqref{eq:staticGFeigenoperators}.

\begin{figure}[!t]
\includegraphics[width=.7\columnwidth]{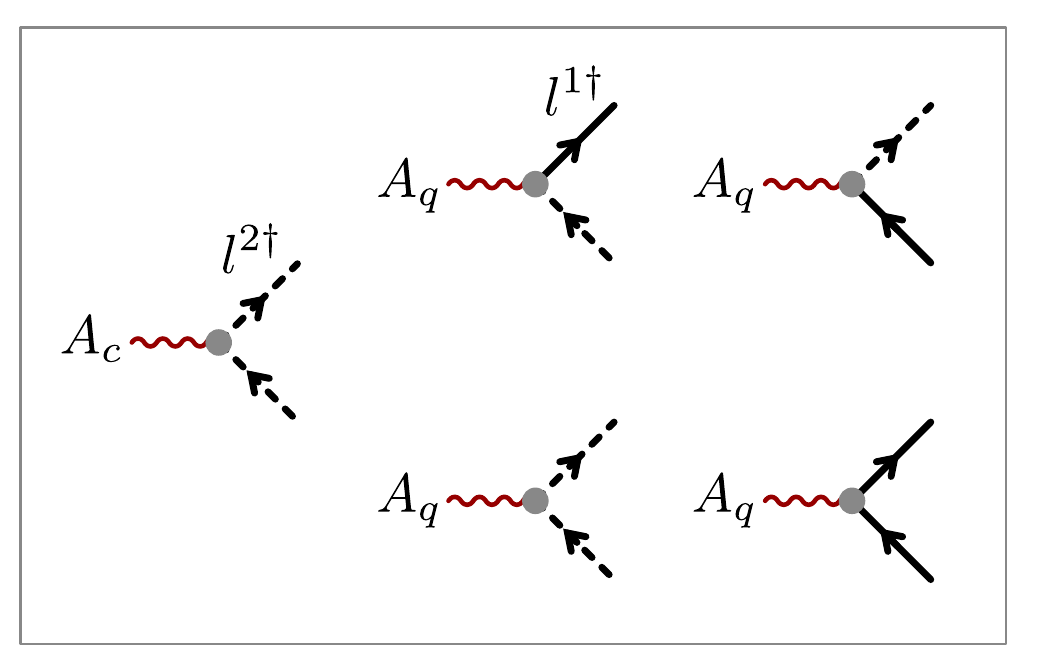}
\caption{Diagrammatic representation of the interaction between gauge fields and effective currents. Continuous lines stand for $l^{1}/l^{2\dagger}$, dashed lines for $l^{2}/l^{1\dagger}$, and wiggly lines for $A_{c/q}$, with Keldysh index clear from the context in this case. }
\label{fig3}
\end{figure}

We move on to the calculation of the connected correlation function $\mean{X^{(1)}_{m.f.}X^{(1)}_{m.f.}}^{\,}_{c.}$, which we illustrate by discussing the components of $\Pi^{IJ}_{ij}(\omega,\bq)$, defined by Eq.~\eqref{eq:defPi}. The first step is to simplify $X^{(1)}_{m.f.}$ by excluding terms contributing only at order $\mc{O}(\bq)$ of the Taylor expansion of $\Pi^{IJ}_{ij}(\omega,\bq)$. This can be done by assuming that all eigenoperators in Eq.~\eqref{eq:MCtermX1} have the same momentum argument, leading e.g. to the replacement 
\begin{align}
l^{\dagger}_{\mb{p}_{1}}(a^{\,}_{i}l)^{\,}_{\mb{p}_{2}} + (a^{\,}_{i\,}l_{\,})^{\dagger}_{\mb{p}_{1}}l^{\,}_{\mb{p}_{2}}\ \to\  2l^{\dagger}_{\mb{p}_{1}}a^{\,}_{i,(\mb{p}_{1}+\mb{p}_{2})/2\,}l^{\,}_{\mb{p}_{2}}\,.
\end{align}
In all the diagrams we compute, momentum conservation actually implies $\mb{p}_{1}=\mb{p}_{2}=(\mb{p}_{1}+\mb{p}_{2})/2$, making the distinction between different field arguments fictitious. The resulting gauge-matter coupling reads:
\begin{align}
\label{eq:MCdecoupl}
X^{(1)}_{m.f.}=\int&2\bar\gamma\, A^{\,}_{c,i}\,l^{1\dagger}_{\,}\big[\sigma^{z},a^{\,}_{i}\big]l^{2}_{\,}\,-\,\bar{\gamma}\,A^{\,}_{q,i}\,\big(l^{2\dagger}_{\,}a_{i}^{\,}l^{1}_{\,}\\
&+l^{2\dagger}_{\,}a^{\,}_{i}\sigma^{z}l^{2}_{\,}-\,l^{1\dagger}_{\,}\sigma^{z}a^{\,}_{i}l^{1}_{\,}-\,l^{1\dagger}_{\,}a_{i}^{\,}l^{2}_{\,}\big).\notag
\end{align}
The corresponding interaction vertices between matter and spatial components of the gauge field are depicted in Fig.~\ref{fig3}. 

The simplest term we consider is $\Pi^{cc}_{ij}(0,\mb{0})$. It involves only $l^{2}$ and $l^{1\dagger}$, and thus vanishes due to $\mean{l^{2}_{\alpha}l_{\beta}^{1\dagger}}=0$. 

\begin{figure}[!]
\includegraphics[width=\columnwidth]{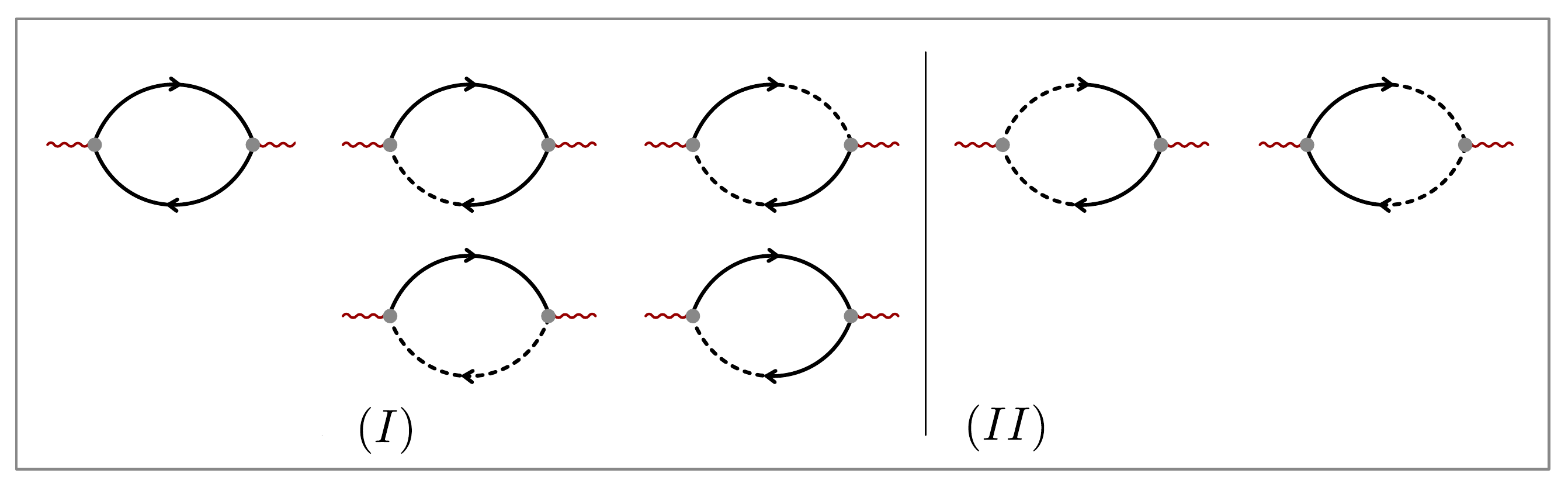}
\caption{Diagrams contributing to $\Pi^{\,}_{qq}(0,\mb{0})$. The two classes of diagrams differ in the matrix structure, see Eqs.~\eqref{eq:Piqq-I} and \eqref{eq:Piqq-II}. }
\label{fig4}
\end{figure}

$\Pi^{qq}_{ij}(0,\mb{0})$ is the sum of seven diagrams, depicted in Fig.~\ref{fig4}. The first five contribute as:
\begin{align}
\label{eq:Piqq-I}
\Pi^{qq}_{ij}(0,\mb{0})\big|_{(I)} =\,& i\bar{\gamma}^{2}\int_{\epsilon,\mb{p}} g^{K}\big( g^{K} - 2 g^{R} + 2 g^{A}\big) \\
&\quad\quad\cdot\text{Tr}\big[\sigma^{z}a_{i,\mb{p}}\sigma^{z}a_{j,\mb{p}}\big]\notag\\
=\,& i\int_{\epsilon,\mb{p}}\frac{4\gamma_{\mb{p}}^{4}}{(\omega^{2}+\gamma^{2}_{\mb{p}})^{2}}\cdot\text{Tr}\big[\sigma^{z}a_{i,\mb{p}}\sigma^{z}a_{j,\mb{p}}\big]\notag\\
=\,&i\bar{\gamma}\int_{\mb{p}}d_{\mb{p}}\,\text{Tr}\big[\sigma^{z}a_{i,\mb{p}}\sigma^{z}a_{j,\mb{p}}\big],\notag
\end{align}
where we omitted the arguments of the Green's functions for brevity, and used $\gamma_{\mb{p}}=\bar{\gamma}d_{\mb{p}}$. The last two diagrams in Fig.~\ref{fig4} contribute instead as:
\begin{align}
\label{eq:Piqq-II}
\Pi^{qq}_{ij}(0,\mb{0})\big|_{(II)}=\,& -i \bar{\gamma}^{2}\int_{\epsilon,\mb{p}} g^{R} g^{A}\,\text{Tr}\big[(a_{i}a_{j})_{\mb{p}}\big]\\
=\,&-i\bar{\gamma}\int_{\mb{p}}d_{\mb{p}}\,\text{Tr}\big[(a_{i}a_{j})_{\mb{p}}\big].\notag
\end{align}
Using Eq.~\eqref{eq:lambda} to identify the parameters $\lambda_{ij}$, the sum of the two parts reads:
\begin{align}
\Pi^{qq}_{ij}(0,\mb{0})\big|_{(I)+(II)} = -i\bar{\gamma}(\lambda_{ij}+\lambda_{ji}),
\end{align}
that cancels the coefficient of the first term in brackets in Eq.~\eqref{eq:X2ev}, after symmetrizing the latter.

\begin{figure}[!t]
\includegraphics[width=\columnwidth]{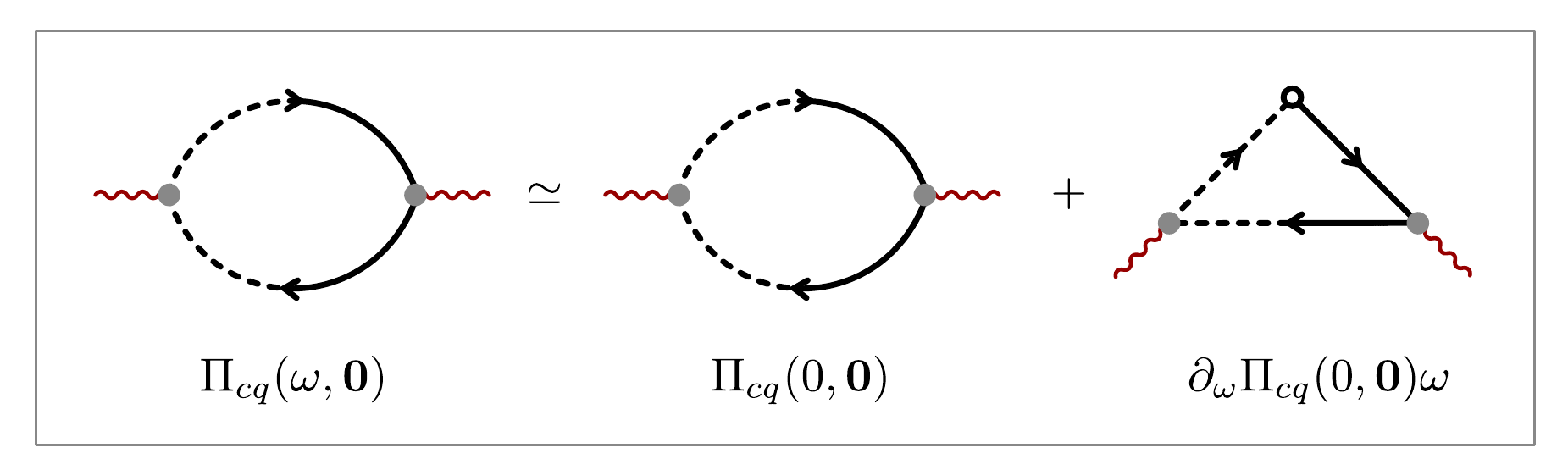}
\caption{Unique diagram contributing to $\Pi^{\,}_{cq}(\omega,\mb{0})$. The Keldysh structure of the triangle diagram can be identified at the first order of the Taylor expansion in $\omega$.} 
\label{fig5}
\end{figure}

The last yet most interesting coefficient is $\Pi^{cq}_{ij}$, as it contains the information on the topological invariant: as shown in Fig.~\ref{fig5}, the triangle diagram of Fig.~\ref{fig2} plus a diverging local contribution both stem from the diagram contributing to $\Pi^{cq}_{ij}(\omega,\mb{0})$ through an expansion in powers of $\omega$. Taking advantage of the simple matrix structure of the Green's functions of the eigenoperators, we get for the full $\Pi^{cq}_{ij}(\omega,\mb{0})$:
\begin{align}
\Pi^{cq}_{ij}(\omega,\mb{0}) =& -2i\bar{\gamma}^{2}\int_{\epsilon,\mb{p}} g^{A}(\epsilon + \omega,\mb{p}) g^{R}(\epsilon,\mb{p})\,\label{eq:Pifull}\\
&\hspace{1.35cm}\cdot \,\text{Tr}\big[\sigma^{z}(a_{i}a_{j} - a_{j}a_{i})_{\mb{p}}\big]=\notag\\
=&\int_{\mb{p}}\frac{\bar{\gamma}^{2}d_{\mb{p}}^{2}}{i\gamma_{\mb{p}}-\omega/2}\,\text{Tr}\big[\sigma^{z}(a_{i}a_{j} - a_{j}a_{i})_{\mb{p}}\big].\notag
\end{align}

Setting $\omega=0$ and expressing Eq.~\eqref{eq:Pifull} in terms of $\lambda_{ij}$, it becomes:
\begin{align}
\label{eq:X1X1ev-local}
\Pi^{cq}_{ij}(0,\mb{0})= i\bar{\gamma}\big(\lambda_{ij}-\lambda_{ji}\big). 
\end{align}
This coefficient cancels the analogous one multiplying the second term in brackets in Eq.~\eqref{eq:X2ev}, upon antisymmetrization of the latter. Moreover, both $\Pi^{qc}_{ij}$ and the coefficient of the third term in Eq.~\eqref{eq:X2ev} are Hermitian conjugates of their $cq$ counterparts, hence they also cancel. The zeroth order of the derivative expansion of $S^{(2)}[A]$ vanishes then exactly, and gauge invariance is preserved as anticipated at the beginning of the section.

At $\mc{O}(\omega)$ we get the topological invariant:
\begin{align}
\label{eq:Pifirst}
\omega\,\partial_{\omega}\Pi^{cq}_{ij}(0,\mb{0}) =\,& -\omega\int_{\mb{p}}\text{Tr}\big[\sigma^{z}(a_{i}a_{j} - a_{j}a_{i})_{\mb{p}}\big]\\
=\,&\frac{i\omega}{2}\,\epsilon_{ij}\,\int_{\mb{p}} \text{Tr}\big[\sigma^{z} F\big] = -i\omega\frac{\mc{\theta}}{2\pi}\epsilon_{ij},\notag
\end{align}
$F= \partial_{q_{1}}a_{2} - \partial_{q_{2}}a_{1}$ being the Berry curvature, $\theta$ the Chern number of the filled band. Manipulations leading to Eq.~\eqref{eq:Pifirst} are shown below. One part of the Chern-Simons action is recovered after substituting Eq.~\eqref{eq:Pifirst} in $S^{(2)}[A]$, namely $\frac{\theta}{4\pi}\int \epsilon_{ij}A^{\,}_{c,i}\partial_{t}A^{\,}_{q,j}$, (partially) proving the relation \eqref{eq:CSFinal} between the CS level and the Chern number. The identification $S^{(2)}[A]=S_{\text{CS}}[A]$ at the first order of the derivative expansion of the gauge action can indeed be confirmed by a more complete yet involved calculation \cite{us}. We remark that such result extends the proof of gauge invariance up to \emph{first} order of the derivative expansion of the gauge action, whereas the previous calculation shows it only at the \emph{zeroth} order.

Let us conclude the section by showing the manipulations leading to the last equality of Eq.~\eqref{eq:Pifirst}. First, the definition $a_{i} = i\partial_{q_{i}}U U^{-1} = -iU \partial_{q_{i}}U^{-1}$ implies that 
\begin{align}
a_{1}a_{2} - a_{2}a_{1} &= \partial_{q_{1}}U\partial_{q_{2}}U^{-1} -\partial_{q_{2}}U\partial_{q_{1}}U^{-1} \\
&= -i(\partial_{q_{1}}a_{2} - \partial_{q_{2}}a_{1}) = iF. \notag
\end{align}
The last equality follows from the zero sum rule obeyed by the Berry curvatures of the bands, i.e., $\int \text{Tr}\, F = 0$ \cite{Asboth2016}.

\subsection{C.\qquad Decay of particle-hole excitations}
\label{app:phdecay}

Here we show that the elementary two-body excitation, the creation of a particle in the upper band and a hole in the lower one, is massive in the strongly interacting model. 

In the first step, we define the states of interest in terms of normalized eigenoperators as
\begin{align}
\label{eq:phexcit}
|\mb{k}\rangle = n^{-1/2}\int_{\mb{q}}\hat{c}^{\dagger}_{1,\mb{q}-\mb{k}}\hat{c}^{\,}_{2,\mb{q}}\,|D\rangle.
\end{align}
They are normalized:
\begin{align}
\langle \mb{k} |\mb{k}'\rangle &=\, n^{-1}\langle D | \int_{\mb{q},\mb{q}'}\hat{c}^{\dagger}_{2,\mb{q}\,}\hat{c}^{\,}_{1,\mb{q}-\mb{k}\,}\hat{c}^{\dagger}_{1,\mb{q}'-\mb{k}'\,}\hat{c}^{\,}_{2,\mb{q}'}\,|D\rangle \notag\\
&=\, n^{-1}\cdot\delta_{\mb{k},\mb{k}'\,}n=\delta_{\mb{k},\mb{k}'}.
\end{align}

As a second step, we define an effective model that captures the essence of the task, by including only the exactly local processes in the generator of dynamics. This way, we take into account the quick decay of the amplitude of the particle-hole excitations but not its slow, subleading dispersion. The new Lindblad operator describes only direct transitions from the upper band to the lower band, local in real space:
\begin{align}
\label{eq:effectiveLindblad}
\hat{L}(\mb{x}) = \hat{c}^{\dagger}_{2}(\mb{x})\hat{c}^{\,}_{1}(\mb{x})& &\Leftrightarrow & &\hat{L}_{\mb{k}} = \int_{\mb{q}} \hat{c}^{\dagger}_{2,\mb{q}-\mb{k}}\hat{c}^{\,}_{1,\mb{q}}.
\end{align}
The damping strength is set to $\gamma m^{2}$, leading to a mean field dissipative gap equal to $\bar{\gamma}m^{2}$ through the same mean field decoupling explained in the main text, equivalent to $\gamma\to\gamma n = \bar{\gamma}$.

The action of the operator \eqref{eq:effectiveLindblad} on the state \eqref{eq:phexcit} is
\begin{align}
\hat{L}_{\mb{p}}|\mb{k}\rangle &=\,n^{-1/2} \int_{\mb{q},\mb{q}'}\hat{c}^{\dagger}_{2,\mb{q}-\mb{p}}\hat{c}^{\,}_{1,\mb{q}}\,\hat{c}^{\dagger}_{1,\mb{q}'-\mb{k}}\hat{c}^{\,}_{2,\mb{q}'}\,|D\rangle\notag\\
&=\, n^{1/2}\delta_{\mb{p},-\mb{k}}|D\rangle.
\end{align}
The anticommutator term in the Liouvillian then yields
\begin{align}
\label{eq:anticommutator-ph}
\int_{\mb{p}}\hat{L}^{\dagger}_{\mb{p}}\hat{L}_{\mb{p}}|\mb{k}\rangle = n|\mb{k}\rangle.
\end{align}
The quantum jump term yields instead
\begin{align}
\label{eq:qjump-ph}
\int_{\mb{p}}\hat{L}_{\mb{p}}|\mb{k}\rangle\langle \mb{k}|\hat{L}^{\dagger}_{\mb{p}} = n|D\rangle \langle D|.
\end{align}

If the initial state is $\hat{\rho}(0) = |\mb{k}\rangle\langle \mb{k}|$, it follows from Eqs.~\eqref{eq:anticommutator-ph} and \eqref{eq:qjump-ph} that an ansatz for the density matrix at all times can be chosen as
\begin{align}
\label{eq:evol-ph}
\hat{\rho} = \rho_{0}|D\rangle\langle D| +\rho_{\mb{k}}|\mb{k}\rangle\langle \mb{k}|\,.
\end{align}
The closed set of equations for $\rho_{0}$ ($0$ particles, $0$ holes) and $\rho_{\mb{k}}$ ($1$ particle, $1$ hole) are:
\begin{align}
\partial_{t}{\rho}_{0} = 2\bar{\gamma}m^{2}\rho_{\mb{k}},& &\partial_{t}{\rho}_{\mb{k}} = -2\bar{\gamma}m^{2}\rho_{\mb{k}}.
\end{align}
Eqs.~\eqref{eq:evol-ph} show that the steady state is approached exponentially fast, i.e., that the elementary two-body excitation is gapped. 

\bibliographystyle{apsrev4-1}
\bibliography{partial-lib}

\begin{thebibliography}{58}%
\makeatletter
\providecommand \@ifxundefined [1]{%
 \@ifx{#1\undefined}
}%
\providecommand \@ifnum [1]{%
 \ifnum #1\expandafter \@firstoftwo
 \else \expandafter \@secondoftwo
 \fi
}%
\providecommand \@ifx [1]{%
 \ifx #1\expandafter \@firstoftwo
 \else \expandafter \@secondoftwo
 \fi
}%
\providecommand \natexlab [1]{#1}%
\providecommand \enquote  [1]{``#1''}%
\providecommand \bibnamefont  [1]{#1}%
\providecommand \bibfnamefont [1]{#1}%
\providecommand \citenamefont [1]{#1}%
\providecommand \href@noop [0]{\@secondoftwo}%
\providecommand \href [0]{\begingroup \@sanitize@url \@href}%
\providecommand \@href[1]{\@@startlink{#1}\@@href}%
\providecommand \@@href[1]{\endgroup#1\@@endlink}%
\providecommand \@sanitize@url [0]{\catcode `\\12\catcode `\$12\catcode
  `\&12\catcode `\#12\catcode `\^12\catcode `\_12\catcode `\%12\relax}%
\providecommand \@@startlink[1]{}%
\providecommand \@@endlink[0]{}%
\providecommand \url  [0]{\begingroup\@sanitize@url \@url }%
\providecommand \@url [1]{\endgroup\@href {#1}{\urlprefix }}%
\providecommand \urlprefix  [0]{URL }%
\providecommand \Eprint [0]{\href }%
\providecommand \doibase [0]{http://dx.doi.org/}%
\providecommand \selectlanguage [0]{\@gobble}%
\providecommand \bibinfo  [0]{\@secondoftwo}%
\providecommand \bibfield  [0]{\@secondoftwo}%
\providecommand \translation [1]{[#1]}%
\providecommand \BibitemOpen [0]{}%
\providecommand \bibitemStop [0]{}%
\providecommand \bibitemNoStop [0]{.\EOS\space}%
\providecommand \EOS [0]{\spacefactor3000\relax}%
\providecommand \BibitemShut  [1]{\csname bibitem#1\endcsname}%
\let\auto@bib@innerbib\@empty
\bibitem [{\citenamefont {Thouless}\ \emph {et~al.}(1982)\citenamefont
  {Thouless}, \citenamefont {Kohmoto}, \citenamefont {Nightingale},\ and\
  \citenamefont {den Nijs}}]{Thouless1982}%
  \BibitemOpen
  \bibfield  {author} {\bibinfo {author} {\bibfnamefont {D.~J.}\ \bibnamefont
  {Thouless}}, \bibinfo {author} {\bibfnamefont {M.}~\bibnamefont {Kohmoto}},
  \bibinfo {author} {\bibfnamefont {M.~P.}\ \bibnamefont {Nightingale}}, \ and\
  \bibinfo {author} {\bibfnamefont {M.}~\bibnamefont {den Nijs}},\ }\href@noop
  {} {\bibfield  {journal} {\bibinfo  {journal} {Phys. Rev. Lett.}\ }\textbf
  {\bibinfo {volume} {49}},\ \bibinfo {pages} {405} (\bibinfo {year}
  {1982})}\BibitemShut {NoStop}%
\bibitem [{\citenamefont {Thouless}(1983)}]{Thouless83}%
  \BibitemOpen
  \bibfield  {author} {\bibinfo {author} {\bibfnamefont {D.~J.}\ \bibnamefont
  {Thouless}},\ }\href {\doibase 10.1103/PhysRevB.27.6083} {\bibfield
  {journal} {\bibinfo  {journal} {Phys. Rev. B}\ }\textbf {\bibinfo {volume}
  {27}},\ \bibinfo {pages} {6083} (\bibinfo {year} {1983})}\BibitemShut
  {NoStop}%
\bibitem [{\citenamefont {Kane}\ and\ \citenamefont {Mele}(2005)}]{KaneMele05}%
  \BibitemOpen
  \bibfield  {author} {\bibinfo {author} {\bibfnamefont {C.~L.}\ \bibnamefont
  {Kane}}\ and\ \bibinfo {author} {\bibfnamefont {E.~J.}\ \bibnamefont
  {Mele}},\ }\href {\doibase 10.1103/PhysRevLett.95.146802} {\bibfield
  {journal} {\bibinfo  {journal} {Phys. Rev. Lett.}\ }\textbf {\bibinfo
  {volume} {95}},\ \bibinfo {pages} {146802} (\bibinfo {year}
  {2005})}\BibitemShut {NoStop}%
\bibitem [{\citenamefont {Hasan}\ and\ \citenamefont
  {Kane}(2010)}]{HasanKane2010}%
  \BibitemOpen
  \bibfield  {author} {\bibinfo {author} {\bibfnamefont {M.~Z.}\ \bibnamefont
  {Hasan}}\ and\ \bibinfo {author} {\bibfnamefont {C.~L.}\ \bibnamefont
  {Kane}},\ }\href {\doibase 10.1103/RevModPhys.82.3045} {\bibfield  {journal}
  {\bibinfo  {journal} {Rev. Mod. Phys.}\ }\textbf {\bibinfo {volume} {82}},\
  \bibinfo {pages} {3045} (\bibinfo {year} {2010})}\BibitemShut {NoStop}%
\bibitem [{\citenamefont {Qi}\ and\ \citenamefont {Zhang}(2011)}]{QiZhang2011}%
  \BibitemOpen
  \bibfield  {author} {\bibinfo {author} {\bibfnamefont {X.-L.}\ \bibnamefont
  {Qi}}\ and\ \bibinfo {author} {\bibfnamefont {S.-C.}\ \bibnamefont {Zhang}},\
  }\href {\doibase 10.1103/RevModPhys.83.1057} {\bibfield  {journal} {\bibinfo
  {journal} {Rev. Mod. Phys.}\ }\textbf {\bibinfo {volume} {83}},\ \bibinfo
  {pages} {1057} (\bibinfo {year} {2011})}\BibitemShut {NoStop}%
\bibitem [{\citenamefont {Asb{\'{o}}th}\ \emph {et~al.}(2016)\citenamefont
  {Asb{\'{o}}th}, \citenamefont {Oroszl{\'{a}}ny},\ and\ \citenamefont
  {P{\'{a}}lyi}}]{Asboth2016}%
  \BibitemOpen
  \bibfield  {author} {\bibinfo {author} {\bibfnamefont {J.~K.}\ \bibnamefont
  {Asb{\'{o}}th}}, \bibinfo {author} {\bibfnamefont {L.}~\bibnamefont
  {Oroszl{\'{a}}ny}}, \ and\ \bibinfo {author} {\bibfnamefont {A.}~\bibnamefont
  {P{\'{a}}lyi}},\ }\href {\doibase 10.1007/978-3-319-25607-8} {\emph {\bibinfo
  {title} {{A Short Course on Topological Insulators}}}},\ \bibinfo {series}
  {Lecture Notes in Physics}, Vol.\ \bibinfo {volume} {919}\ (\bibinfo
  {publisher} {Springer International Publishing},\ \bibinfo {address} {Cham},\
  \bibinfo {year} {2016})\BibitemShut {NoStop}%
\bibitem [{\citenamefont {Lindner}\ \emph {et~al.}(2011)\citenamefont
  {Lindner}, \citenamefont {Refael},\ and\ \citenamefont
  {Galitski}}]{Lindner2011}%
  \BibitemOpen
  \bibfield  {author} {\bibinfo {author} {\bibfnamefont {N.~H.}\ \bibnamefont
  {Lindner}}, \bibinfo {author} {\bibfnamefont {G.}~\bibnamefont {Refael}}, \
  and\ \bibinfo {author} {\bibfnamefont {V.}~\bibnamefont {Galitski}},\ }\href
  {\doibase 10.1038/nphys1926} {\bibfield  {journal} {\bibinfo  {journal} {Nat.
  Phys.}\ }\textbf {\bibinfo {volume} {7}},\ \bibinfo {pages} {490} (\bibinfo
  {year} {2011})}\BibitemShut {NoStop}%
\bibitem [{\citenamefont {Rudner}\ \emph {et~al.}(2013)\citenamefont {Rudner},
  \citenamefont {Lindner}, \citenamefont {Berg},\ and\ \citenamefont
  {Levin}}]{Rudner2013}%
  \BibitemOpen
  \bibfield  {author} {\bibinfo {author} {\bibfnamefont {M.~S.}\ \bibnamefont
  {Rudner}}, \bibinfo {author} {\bibfnamefont {N.~H.}\ \bibnamefont {Lindner}},
  \bibinfo {author} {\bibfnamefont {E.}~\bibnamefont {Berg}}, \ and\ \bibinfo
  {author} {\bibfnamefont {M.}~\bibnamefont {Levin}},\ }\href {\doibase
  10.1103/PhysRevX.3.031005} {\bibfield  {journal} {\bibinfo  {journal} {Phys.
  Rev. X}\ }\textbf {\bibinfo {volume} {3}},\ \bibinfo {pages} {031005}
  (\bibinfo {year} {2013})}\BibitemShut {NoStop}%
\bibitem [{\citenamefont {{Rudner}}\ and\ \citenamefont
  {{Lindner}}()}]{Rudner2019}%
  \BibitemOpen
  \bibfield  {author} {\bibinfo {author} {\bibfnamefont {M.~S.}\ \bibnamefont
  {{Rudner}}}\ and\ \bibinfo {author} {\bibfnamefont {N.~H.}\ \bibnamefont
  {{Lindner}}},\ }\href@noop {} {\ }\Eprint
  {http://arxiv.org/abs/arXiv:1909.02008 (2019)} {arXiv:1909.02008 (2019)}
  \BibitemShut {NoStop}%
\bibitem [{\citenamefont {Diehl}\ \emph {et~al.}(2011)\citenamefont {Diehl},
  \citenamefont {Rico}, \citenamefont {Baranov},\ and\ \citenamefont
  {Zoller}}]{Diehl2011}%
  \BibitemOpen
  \bibfield  {author} {\bibinfo {author} {\bibfnamefont {S.}~\bibnamefont
  {Diehl}}, \bibinfo {author} {\bibfnamefont {E.}~\bibnamefont {Rico}},
  \bibinfo {author} {\bibfnamefont {M.~A.}\ \bibnamefont {Baranov}}, \ and\
  \bibinfo {author} {\bibfnamefont {P.}~\bibnamefont {Zoller}},\ }\href
  {\doibase 10.1038/nphys2106} {\bibfield  {journal} {\bibinfo  {journal} {Nat.
  Phys.}\ }\textbf {\bibinfo {volume} {7}},\ \bibinfo {pages} {971} (\bibinfo
  {year} {2011})}\BibitemShut {NoStop}%
\bibitem [{\citenamefont {Bardyn}\ \emph {et~al.}(2013)\citenamefont {Bardyn},
  \citenamefont {Baranov}, \citenamefont {Kraus}, \citenamefont {Rico},
  \citenamefont {\.{I}mamo\u{g}lu}, \citenamefont {Zoller},\ and\ \citenamefont
  {Diehl}}]{Bardyn2013}%
  \BibitemOpen
  \bibfield  {author} {\bibinfo {author} {\bibfnamefont {C.-E.}\ \bibnamefont
  {Bardyn}}, \bibinfo {author} {\bibfnamefont {M.~A.}\ \bibnamefont {Baranov}},
  \bibinfo {author} {\bibfnamefont {C.~V.}\ \bibnamefont {Kraus}}, \bibinfo
  {author} {\bibfnamefont {E.}~\bibnamefont {Rico}}, \bibinfo {author}
  {\bibfnamefont {A.}~\bibnamefont {\.{I}mamo\u{g}lu}}, \bibinfo {author}
  {\bibfnamefont {P.}~\bibnamefont {Zoller}}, \ and\ \bibinfo {author}
  {\bibfnamefont {S.}~\bibnamefont {Diehl}},\ }\href {\doibase
  10.1088/1367-2630/15/8/085001} {\bibfield  {journal} {\bibinfo  {journal}
  {New J. Phys.}\ }\textbf {\bibinfo {volume} {15}},\ \bibinfo {pages} {085001}
  (\bibinfo {year} {2013})}\BibitemShut {NoStop}%
\bibitem [{\citenamefont {Budich}\ \emph {et~al.}(2015)\citenamefont {Budich},
  \citenamefont {Zoller},\ and\ \citenamefont {Diehl}}]{Budich2015}%
  \BibitemOpen
  \bibfield  {author} {\bibinfo {author} {\bibfnamefont {J.~C.}\ \bibnamefont
  {Budich}}, \bibinfo {author} {\bibfnamefont {P.}~\bibnamefont {Zoller}}, \
  and\ \bibinfo {author} {\bibfnamefont {S.}~\bibnamefont {Diehl}},\ }\href
  {\doibase 10.1103/PhysRevA.91.042117} {\bibfield  {journal} {\bibinfo
  {journal} {Phys. Rev. A}\ }\textbf {\bibinfo {volume} {91}},\ \bibinfo
  {pages} {042117} (\bibinfo {year} {2015})}\BibitemShut {NoStop}%
\bibitem [{\citenamefont {Weimann}\ \emph {et~al.}(2017)\citenamefont
  {Weimann}, \citenamefont {Kremer}, \citenamefont {Plotnik}, \citenamefont
  {Lumer}, \citenamefont {Nolte}, \citenamefont {Makris}, \citenamefont
  {Segev}, \citenamefont {Rechtsman},\ and\ \citenamefont
  {Szameit}}]{Weimann2017}%
  \BibitemOpen
  \bibfield  {author} {\bibinfo {author} {\bibfnamefont {S.}~\bibnamefont
  {Weimann}}, \bibinfo {author} {\bibfnamefont {M.}~\bibnamefont {Kremer}},
  \bibinfo {author} {\bibfnamefont {Y.}~\bibnamefont {Plotnik}}, \bibinfo
  {author} {\bibfnamefont {Y.}~\bibnamefont {Lumer}}, \bibinfo {author}
  {\bibfnamefont {S.}~\bibnamefont {Nolte}}, \bibinfo {author} {\bibfnamefont
  {K.~G.}\ \bibnamefont {Makris}}, \bibinfo {author} {\bibfnamefont
  {M.}~\bibnamefont {Segev}}, \bibinfo {author} {\bibfnamefont {M.~C.}\
  \bibnamefont {Rechtsman}}, \ and\ \bibinfo {author} {\bibfnamefont
  {A.}~\bibnamefont {Szameit}},\ }\href {\doibase 10.1038/nmat4811} {\bibfield
  {journal} {\bibinfo  {journal} {Nat. Materials}\ }\textbf {\bibinfo {volume}
  {16}},\ \bibinfo {pages} {433} (\bibinfo {year} {2017})}\BibitemShut
  {NoStop}%
\bibitem [{\citenamefont {Zhou}\ \emph {et~al.}(2018)\citenamefont {Zhou},
  \citenamefont {Peng}, \citenamefont {Yoon}, \citenamefont {Hsu},
  \citenamefont {Nelson}, \citenamefont {Fu}, \citenamefont {Joannopoulos},
  \citenamefont {Solja{\v{c}}i{\'{c}}},\ and\ \citenamefont {Zhen}}]{Zhou2018}%
  \BibitemOpen
  \bibfield  {author} {\bibinfo {author} {\bibfnamefont {H.}~\bibnamefont
  {Zhou}}, \bibinfo {author} {\bibfnamefont {C.}~\bibnamefont {Peng}}, \bibinfo
  {author} {\bibfnamefont {Y.}~\bibnamefont {Yoon}}, \bibinfo {author}
  {\bibfnamefont {C.~W.}\ \bibnamefont {Hsu}}, \bibinfo {author} {\bibfnamefont
  {K.~A.}\ \bibnamefont {Nelson}}, \bibinfo {author} {\bibfnamefont
  {L.}~\bibnamefont {Fu}}, \bibinfo {author} {\bibfnamefont {J.~D.}\
  \bibnamefont {Joannopoulos}}, \bibinfo {author} {\bibfnamefont
  {M.}~\bibnamefont {Solja{\v{c}}i{\'{c}}}}, \ and\ \bibinfo {author}
  {\bibfnamefont {B.}~\bibnamefont {Zhen}},\ }\href {\doibase
  10.1126/science.aap9859} {\bibfield  {journal} {\bibinfo  {journal} {Science
  (New York, N.Y.)}\ }\textbf {\bibinfo {volume} {359}},\ \bibinfo {pages}
  {1009} (\bibinfo {year} {2018})}\BibitemShut {NoStop}%
\bibitem [{\citenamefont {Gong}\ \emph {et~al.}(2018)\citenamefont {Gong},
  \citenamefont {Ashida}, \citenamefont {Kawabata}, \citenamefont {Takasan},
  \citenamefont {Higashikawa},\ and\ \citenamefont {Ueda}}]{Gong2018}%
  \BibitemOpen
  \bibfield  {author} {\bibinfo {author} {\bibfnamefont {Z.}~\bibnamefont
  {Gong}}, \bibinfo {author} {\bibfnamefont {Y.}~\bibnamefont {Ashida}},
  \bibinfo {author} {\bibfnamefont {K.}~\bibnamefont {Kawabata}}, \bibinfo
  {author} {\bibfnamefont {K.}~\bibnamefont {Takasan}}, \bibinfo {author}
  {\bibfnamefont {S.}~\bibnamefont {Higashikawa}}, \ and\ \bibinfo {author}
  {\bibfnamefont {M.}~\bibnamefont {Ueda}},\ }\href {\doibase
  10.1103/PhysRevX.8.031079} {\bibfield  {journal} {\bibinfo  {journal} {Phys.
  Rev. X}\ }\textbf {\bibinfo {volume} {8}},\ \bibinfo {pages} {031079}
  (\bibinfo {year} {2018})}\BibitemShut {NoStop}%
\bibitem [{\citenamefont {Kunst}\ \emph {et~al.}(2018)\citenamefont {Kunst},
  \citenamefont {Edvardsson}, \citenamefont {Budich},\ and\ \citenamefont
  {Bergholtz}}]{Kunst2018}%
  \BibitemOpen
  \bibfield  {author} {\bibinfo {author} {\bibfnamefont {F.~K.}\ \bibnamefont
  {Kunst}}, \bibinfo {author} {\bibfnamefont {E.}~\bibnamefont {Edvardsson}},
  \bibinfo {author} {\bibfnamefont {J.~C.}\ \bibnamefont {Budich}}, \ and\
  \bibinfo {author} {\bibfnamefont {E.~J.}\ \bibnamefont {Bergholtz}},\ }\href
  {\doibase 10.1103/PhysRevLett.121.026808} {\bibfield  {journal} {\bibinfo
  {journal} {Phys. Rev. Lett.}\ }\textbf {\bibinfo {volume} {121}},\ \bibinfo
  {pages} {026808} (\bibinfo {year} {2018})}\BibitemShut {NoStop}%
\bibitem [{\citenamefont {Kawabata}\ \emph {et~al.}(2019)\citenamefont
  {Kawabata}, \citenamefont {Shiozaki}, \citenamefont {Ueda},\ and\
  \citenamefont {Sato}}]{Kawabata2019}%
  \BibitemOpen
  \bibfield  {author} {\bibinfo {author} {\bibfnamefont {K.}~\bibnamefont
  {Kawabata}}, \bibinfo {author} {\bibfnamefont {K.}~\bibnamefont {Shiozaki}},
  \bibinfo {author} {\bibfnamefont {M.}~\bibnamefont {Ueda}}, \ and\ \bibinfo
  {author} {\bibfnamefont {M.}~\bibnamefont {Sato}},\ }\href {\doibase
  10.1103/PhysRevX.9.041015} {\bibfield  {journal} {\bibinfo  {journal} {Phys.
  Rev. X}\ }\textbf {\bibinfo {volume} {9}},\ \bibinfo {pages} {041015}
  (\bibinfo {year} {2019})}\BibitemShut {NoStop}%
\bibitem [{\citenamefont {Goldman}\ \emph {et~al.}(2016)\citenamefont
  {Goldman}, \citenamefont {Budich},\ and\ \citenamefont
  {Zoller}}]{Goldman2016}%
  \BibitemOpen
  \bibfield  {author} {\bibinfo {author} {\bibfnamefont {N.}~\bibnamefont
  {Goldman}}, \bibinfo {author} {\bibfnamefont {J.~C.}\ \bibnamefont {Budich}},
  \ and\ \bibinfo {author} {\bibfnamefont {P.}~\bibnamefont {Zoller}},\ }\href
  {\doibase 10.1038/nphys3803} {\bibfield  {journal} {\bibinfo  {journal} {Nat.
  Phys.}\ }\textbf {\bibinfo {volume} {12}},\ \bibinfo {pages} {639} (\bibinfo
  {year} {2016})}\BibitemShut {NoStop}%
\bibitem [{\citenamefont {Lu}\ \emph {et~al.}(2014)\citenamefont {Lu},
  \citenamefont {Joannopoulos},\ and\ \citenamefont
  {Solja{\v{c}}i{\'{c}}}}]{Lu2014}%
  \BibitemOpen
  \bibfield  {author} {\bibinfo {author} {\bibfnamefont {L.}~\bibnamefont
  {Lu}}, \bibinfo {author} {\bibfnamefont {J.~D.}\ \bibnamefont
  {Joannopoulos}}, \ and\ \bibinfo {author} {\bibfnamefont {M.}~\bibnamefont
  {Solja{\v{c}}i{\'{c}}}},\ }\href {\doibase 10.1038/nphoton.2014.248}
  {\bibfield  {journal} {\bibinfo  {journal} {Nat. Photon.}\ }\textbf {\bibinfo
  {volume} {8}},\ \bibinfo {pages} {821} (\bibinfo {year} {2014})}\BibitemShut
  {NoStop}%
\bibitem [{\citenamefont {Ozawa}\ \emph {et~al.}(2019)\citenamefont {Ozawa},
  \citenamefont {Price}, \citenamefont {Amo}, \citenamefont {Goldman},
  \citenamefont {Hafezi}, \citenamefont {Lu}, \citenamefont {Rechtsman},
  \citenamefont {Schuster}, \citenamefont {Simon}, \citenamefont {Zilberberg},\
  and\ \citenamefont {Carusotto}}]{Ozawa2019}%
  \BibitemOpen
  \bibfield  {author} {\bibinfo {author} {\bibfnamefont {T.}~\bibnamefont
  {Ozawa}}, \bibinfo {author} {\bibfnamefont {H.~M.}\ \bibnamefont {Price}},
  \bibinfo {author} {\bibfnamefont {A.}~\bibnamefont {Amo}}, \bibinfo {author}
  {\bibfnamefont {N.}~\bibnamefont {Goldman}}, \bibinfo {author} {\bibfnamefont
  {M.}~\bibnamefont {Hafezi}}, \bibinfo {author} {\bibfnamefont
  {L.}~\bibnamefont {Lu}}, \bibinfo {author} {\bibfnamefont {M.~C.}\
  \bibnamefont {Rechtsman}}, \bibinfo {author} {\bibfnamefont {D.}~\bibnamefont
  {Schuster}}, \bibinfo {author} {\bibfnamefont {J.}~\bibnamefont {Simon}},
  \bibinfo {author} {\bibfnamefont {O.}~\bibnamefont {Zilberberg}}, \ and\
  \bibinfo {author} {\bibfnamefont {I.}~\bibnamefont {Carusotto}},\ }\href
  {\doibase 10.1103/RevModPhys.91.015006} {\bibfield  {journal} {\bibinfo
  {journal} {Rev. Mod. Phys.}\ }\textbf {\bibinfo {volume} {91}},\ \bibinfo
  {pages} {015006} (\bibinfo {year} {2019})}\BibitemShut {NoStop}%
\bibitem [{\citenamefont {St-Jean}\ \emph {et~al.}(2017)\citenamefont
  {St-Jean}, \citenamefont {Goblot}, \citenamefont {Galopin}, \citenamefont
  {Lema{\^{i}}tre}, \citenamefont {Ozawa}, \citenamefont {{Le Gratiet}},
  \citenamefont {Sagnes}, \citenamefont {Bloch},\ and\ \citenamefont
  {Amo}}]{St-Jean2017}%
  \BibitemOpen
  \bibfield  {author} {\bibinfo {author} {\bibfnamefont {P.}~\bibnamefont
  {St-Jean}}, \bibinfo {author} {\bibfnamefont {V.}~\bibnamefont {Goblot}},
  \bibinfo {author} {\bibfnamefont {E.}~\bibnamefont {Galopin}}, \bibinfo
  {author} {\bibfnamefont {A.}~\bibnamefont {Lema{\^{i}}tre}}, \bibinfo
  {author} {\bibfnamefont {T.}~\bibnamefont {Ozawa}}, \bibinfo {author}
  {\bibfnamefont {L.}~\bibnamefont {{Le Gratiet}}}, \bibinfo {author}
  {\bibfnamefont {I.}~\bibnamefont {Sagnes}}, \bibinfo {author} {\bibfnamefont
  {J.}~\bibnamefont {Bloch}}, \ and\ \bibinfo {author} {\bibfnamefont
  {A.}~\bibnamefont {Amo}},\ }\href {\doibase 10.1038/s41566-017-0006-2}
  {\bibfield  {journal} {\bibinfo  {journal} {Nat. Photon.}\ }\textbf {\bibinfo
  {volume} {11}},\ \bibinfo {pages} {651} (\bibinfo {year} {2017})}\BibitemShut
  {NoStop}%
\bibitem [{\citenamefont {Klembt}\ \emph {et~al.}(2018)\citenamefont {Klembt},
  \citenamefont {Harder}, \citenamefont {Egorov}, \citenamefont {Winkler},
  \citenamefont {Ge}, \citenamefont {Bandres}, \citenamefont {Emmerling},
  \citenamefont {Worschech}, \citenamefont {Liew}, \citenamefont {Segev},
  \citenamefont {Schneider},\ and\ \citenamefont {H{\"{o}}fling}}]{Klembt2018}%
  \BibitemOpen
  \bibfield  {author} {\bibinfo {author} {\bibfnamefont {S.}~\bibnamefont
  {Klembt}}, \bibinfo {author} {\bibfnamefont {T.~H.}\ \bibnamefont {Harder}},
  \bibinfo {author} {\bibfnamefont {O.~A.}\ \bibnamefont {Egorov}}, \bibinfo
  {author} {\bibfnamefont {K.}~\bibnamefont {Winkler}}, \bibinfo {author}
  {\bibfnamefont {R.}~\bibnamefont {Ge}}, \bibinfo {author} {\bibfnamefont
  {M.~A.}\ \bibnamefont {Bandres}}, \bibinfo {author} {\bibfnamefont
  {M.}~\bibnamefont {Emmerling}}, \bibinfo {author} {\bibfnamefont
  {L.}~\bibnamefont {Worschech}}, \bibinfo {author} {\bibfnamefont {T.~C.~H.}\
  \bibnamefont {Liew}}, \bibinfo {author} {\bibfnamefont {M.}~\bibnamefont
  {Segev}}, \bibinfo {author} {\bibfnamefont {C.}~\bibnamefont {Schneider}}, \
  and\ \bibinfo {author} {\bibfnamefont {S.}~\bibnamefont {H{\"{o}}fling}},\
  }\href {\doibase 10.1038/s41586-018-0601-5} {\bibfield  {journal} {\bibinfo
  {journal} {Nature}\ }\textbf {\bibinfo {volume} {562}},\ \bibinfo {pages}
  {552} (\bibinfo {year} {2018})}\BibitemShut {NoStop}%
\bibitem [{\citenamefont {Zhang}\ \emph {et~al.}(1989)\citenamefont {Zhang},
  \citenamefont {Hansson},\ and\ \citenamefont {Kivelson}}]{Zhang1989}%
  \BibitemOpen
  \bibfield  {author} {\bibinfo {author} {\bibfnamefont {S.~C.}\ \bibnamefont
  {Zhang}}, \bibinfo {author} {\bibfnamefont {T.~H.}\ \bibnamefont {Hansson}},
  \ and\ \bibinfo {author} {\bibfnamefont {S.}~\bibnamefont {Kivelson}},\
  }\href {\doibase 10.1103/PhysRevLett.62.82} {\bibfield  {journal} {\bibinfo
  {journal} {Phys. Rev. Lett.}\ }\textbf {\bibinfo {volume} {62}},\ \bibinfo
  {pages} {82} (\bibinfo {year} {1989})}\BibitemShut {NoStop}%
\bibitem [{\citenamefont {Lopez}\ and\ \citenamefont
  {Fradkin}(1991)}]{Lopez1991}%
  \BibitemOpen
  \bibfield  {author} {\bibinfo {author} {\bibfnamefont {A.}~\bibnamefont
  {Lopez}}\ and\ \bibinfo {author} {\bibfnamefont {E.}~\bibnamefont
  {Fradkin}},\ }\href {\doibase 10.1103/PhysRevB.44.5246} {\bibfield  {journal}
  {\bibinfo  {journal} {Phys. Rev. B}\ }\textbf {\bibinfo {volume} {44}},\
  \bibinfo {pages} {5246} (\bibinfo {year} {1991})}\BibitemShut {NoStop}%
\bibitem [{\citenamefont {Kou}\ \emph {et~al.}(2008{\natexlab{a}})\citenamefont
  {Kou}, \citenamefont {Levin},\ and\ \citenamefont {Wen}}]{LevinWen08}%
  \BibitemOpen
  \bibfield  {author} {\bibinfo {author} {\bibfnamefont {S.-P.}\ \bibnamefont
  {Kou}}, \bibinfo {author} {\bibfnamefont {M.}~\bibnamefont {Levin}}, \ and\
  \bibinfo {author} {\bibfnamefont {X.-G.}\ \bibnamefont {Wen}},\ }\href
  {\doibase 10.1103/PhysRevB.78.155134} {\bibfield  {journal} {\bibinfo
  {journal} {Phys. Rev. B}\ }\textbf {\bibinfo {volume} {78}},\ \bibinfo
  {pages} {155134} (\bibinfo {year} {2008}{\natexlab{a}})}\BibitemShut
  {NoStop}%
\bibitem [{\citenamefont {Qi}\ \emph {et~al.}(2006)\citenamefont {Qi},
  \citenamefont {Wu},\ and\ \citenamefont {Zhang}}]{Qi2008}%
  \BibitemOpen
  \bibfield  {author} {\bibinfo {author} {\bibfnamefont {X.-L.}\ \bibnamefont
  {Qi}}, \bibinfo {author} {\bibfnamefont {Y.-S.}\ \bibnamefont {Wu}}, \ and\
  \bibinfo {author} {\bibfnamefont {S.-C.}\ \bibnamefont {Zhang}},\ }\href
  {\doibase 10.1103/PhysRevB.74.085308} {\bibfield  {journal} {\bibinfo
  {journal} {Phys. Rev. B}\ }\textbf {\bibinfo {volume} {74}},\ \bibinfo
  {pages} {085308} (\bibinfo {year} {2006})}\BibitemShut {NoStop}%
\bibitem [{\citenamefont {Redlich}(1984)}]{Redlich1984}%
  \BibitemOpen
  \bibfield  {author} {\bibinfo {author} {\bibfnamefont {A.~N.}\ \bibnamefont
  {Redlich}},\ }\href {\doibase 10.1103/PhysRevD.29.2366} {\bibfield  {journal}
  {\bibinfo  {journal} {Phys. Rev. D}\ }\textbf {\bibinfo {volume} {29}},\
  \bibinfo {pages} {2366} (\bibinfo {year} {1984})}\BibitemShut {NoStop}%
\bibitem [{\citenamefont {Ryu}\ \emph {et~al.}(2012)\citenamefont {Ryu},
  \citenamefont {Moore},\ and\ \citenamefont {Ludwig}}]{Ryu2012}%
  \BibitemOpen
  \bibfield  {author} {\bibinfo {author} {\bibfnamefont {S.}~\bibnamefont
  {Ryu}}, \bibinfo {author} {\bibfnamefont {J.~E.}\ \bibnamefont {Moore}}, \
  and\ \bibinfo {author} {\bibfnamefont {A.~W.~W.}\ \bibnamefont {Ludwig}},\
  }\href {\doibase 10.1103/PhysRevB.85.045104} {\bibfield  {journal} {\bibinfo
  {journal} {Phys. Rev. B}\ }\textbf {\bibinfo {volume} {85}},\ \bibinfo
  {pages} {045104} (\bibinfo {year} {2012})}\BibitemShut {NoStop}%
\bibitem [{\citenamefont {Altland}\ and\ \citenamefont
  {Bagrets}(2016)}]{Bagrets16}%
  \BibitemOpen
  \bibfield  {author} {\bibinfo {author} {\bibfnamefont {A.}~\bibnamefont
  {Altland}}\ and\ \bibinfo {author} {\bibfnamefont {D.}~\bibnamefont
  {Bagrets}},\ }\href {\doibase 10.1103/PhysRevB.93.075113} {\bibfield
  {journal} {\bibinfo  {journal} {Phys. Rev. B}\ }\textbf {\bibinfo {volume}
  {93}},\ \bibinfo {pages} {075113} (\bibinfo {year} {2016})}\BibitemShut
  {NoStop}%
\bibitem [{\citenamefont {Wen}(1995)}]{Wen1995}%
  \BibitemOpen
  \bibfield  {author} {\bibinfo {author} {\bibfnamefont {X.~G.}\ \bibnamefont
  {Wen}},\ }\href@noop {} {\bibfield  {journal} {\bibinfo  {journal} {Adv.
  Phys.}\ }\textbf {\bibinfo {volume} {44}},\ \bibinfo {pages} {405} (\bibinfo
  {year} {1995})}\BibitemShut {NoStop}%
\bibitem [{\citenamefont {Tong}()}]{Tong2016}%
  \BibitemOpen
  \bibfield  {author} {\bibinfo {author} {\bibfnamefont {D.}~\bibnamefont
  {Tong}},\ }\href@noop {} {\ }\Eprint {http://arxiv.org/abs/arXiv:1606.06687
  (2016)} {arXiv:1606.06687 (2016)} \BibitemShut {NoStop}%
\bibitem [{\citenamefont {Kou}\ \emph {et~al.}(2008{\natexlab{b}})\citenamefont
  {Kou}, \citenamefont {Levin},\ and\ \citenamefont {Wen}}]{Wen2008}%
  \BibitemOpen
  \bibfield  {author} {\bibinfo {author} {\bibfnamefont {S.-P.}\ \bibnamefont
  {Kou}}, \bibinfo {author} {\bibfnamefont {M.}~\bibnamefont {Levin}}, \ and\
  \bibinfo {author} {\bibfnamefont {X.-G.}\ \bibnamefont {Wen}},\ }\href
  {\doibase 10.1103/PhysRevB.78.155134} {\bibfield  {journal} {\bibinfo
  {journal} {Phys. Rev. B}\ }\textbf {\bibinfo {volume} {78}},\ \bibinfo
  {pages} {155134} (\bibinfo {year} {2008}{\natexlab{b}})}\BibitemShut
  {NoStop}%
\bibitem [{\citenamefont {{Goldstein}}()}]{goldstein2018}%
  \BibitemOpen
  \bibfield  {author} {\bibinfo {author} {\bibfnamefont {M.}~\bibnamefont
  {{Goldstein}}},\ }\href@noop {} {\ }\Eprint
  {http://arxiv.org/abs/arXiv:1810.12050 (2018)} {arXiv:1810.12050 (2018)}
  \BibitemShut {NoStop}%
\bibitem [{\citenamefont {{Shavit}}\ and\ \citenamefont
  {{Goldstein}}()}]{shavit2019}%
  \BibitemOpen
  \bibfield  {author} {\bibinfo {author} {\bibfnamefont {G.}~\bibnamefont
  {{Shavit}}}\ and\ \bibinfo {author} {\bibfnamefont {M.}~\bibnamefont
  {{Goldstein}}},\ }\href@noop {} {\ }\Eprint
  {http://arxiv.org/abs/arXiv:1903.05336 (2019)} {arXiv:1903.05336 (2019)}
  \BibitemShut {NoStop}%
\bibitem [{\citenamefont {Lindblad}(1976)}]{Lindblad1976}%
  \BibitemOpen
  \bibfield  {author} {\bibinfo {author} {\bibfnamefont {G.}~\bibnamefont
  {Lindblad}},\ }\href {https://projecteuclid.org:443/euclid.cmp/1103899849}
  {\bibfield  {journal} {\bibinfo  {journal} {Comm. Math. Phys.}\ }\textbf
  {\bibinfo {volume} {48}},\ \bibinfo {pages} {119} (\bibinfo {year}
  {1976})}\BibitemShut {NoStop}%
\bibitem [{\citenamefont {Breuer}\ \emph {et~al.}(2002)\citenamefont {Breuer},
  \citenamefont {Petruccione},\ and\ \citenamefont {Petruccione}}]{breuer2002}%
  \BibitemOpen
  \bibfield  {author} {\bibinfo {author} {\bibfnamefont {H.}~\bibnamefont
  {Breuer}}, \bibinfo {author} {\bibfnamefont {F.}~\bibnamefont {Petruccione}},
  \ and\ \bibinfo {author} {\bibfnamefont {S.}~\bibnamefont {Petruccione}},\
  }\href {https://books.google.de/books?id=0Yx5VzaMYm8C} {\emph {\bibinfo
  {title} {The Theory of Open Quantum Systems}}}\ (\bibinfo  {publisher}
  {Oxford University Press},\ \bibinfo {year} {2002})\BibitemShut {NoStop}%
\bibitem [{\citenamefont {Laughlin}(1981)}]{Laughlin81}%
  \BibitemOpen
  \bibfield  {author} {\bibinfo {author} {\bibfnamefont {R.~B.}\ \bibnamefont
  {Laughlin}},\ }\href {\doibase 10.1103/PhysRevB.23.5632} {\bibfield
  {journal} {\bibinfo  {journal} {Phys. Rev. B}\ }\textbf {\bibinfo {volume}
  {23}},\ \bibinfo {pages} {5632} (\bibinfo {year} {1981})}\BibitemShut
  {NoStop}%
\bibitem [{Note1()}]{Note1}%
  \BibitemOpen
  \bibinfo {note} {Einstein's summation convention on repeated indices is
  assumed unless otherwise specified}\BibitemShut {NoStop}%
\bibitem [{\citenamefont {Kamenev}(2011)}]{Kamenev2011b}%
  \BibitemOpen
  \bibfield  {author} {\bibinfo {author} {\bibfnamefont {A.}~\bibnamefont
  {Kamenev}},\ }\href {\doibase 10.1007/s13398-014-0173-7.2} {\emph {\bibinfo
  {title} {{Field theory of non-equilibrium systems}}}}\ (\bibinfo  {publisher}
  {Cambridge University Press},\ \bibinfo {year} {2011})\BibitemShut {NoStop}%
\bibitem [{\citenamefont {Sieberer}\ \emph {et~al.}(2016)\citenamefont
  {Sieberer}, \citenamefont {Buchhold},\ and\ \citenamefont
  {Diehl}}]{Sieberer2015}%
  \BibitemOpen
  \bibfield  {author} {\bibinfo {author} {\bibfnamefont {L.~M.}\ \bibnamefont
  {Sieberer}}, \bibinfo {author} {\bibfnamefont {M.}~\bibnamefont {Buchhold}},
  \ and\ \bibinfo {author} {\bibfnamefont {S.}~\bibnamefont {Diehl}},\
  }\href@noop {} {\bibfield  {journal} {\bibinfo  {journal} {Rep. Prog. Phys}\
  }\textbf {\bibinfo {volume} {79}} (\bibinfo {year} {2016})}\BibitemShut
  {NoStop}%
\bibitem [{Note20()}]{Note20}%
  \BibitemOpen
  \bibinfo {note} {See Supplemental Material for details on the self-consistent
  Born approximation, on the evaluation of the prefactor of the Chern-Simons
  action, and a discussion on the existence of a many-body dissipative gap
  through a paradigmatic example}\BibitemShut {NoStop}%
\bibitem [{Note21()}]{Note21}%
  \BibitemOpen
  \bibinfo {note} {In terms of contour fields, suppressing all arguments but
  time for simplicity, they are equal to {$G^{R}(t) = \theta
  (t)(G^{>}(t)-G^{<}(t))$} and {$G^{K}(t) = G^{>}(t)-G^{<}(t)$}, with
  {$iG^{>/<}_{\alpha \beta }(t) = \delimiter "426830A \psi _{\mp ,\alpha
  }^{\protect \tmspace +\thinmuskip {.1667em}}(t)\psi _{\pm ,\beta }^{\dagger
  }(0) \delimiter "526930B $} \cite {Kamenev2011b}\label
  {Keldyshdefinition}}\BibitemShut {NoStop}%
\bibitem [{\citenamefont {Avron}\ \emph {et~al.}(2011)\citenamefont {Avron},
  \citenamefont {Fraas}, \citenamefont {Graf},\ and\ \citenamefont
  {Kenneth}}]{Avron2011}%
  \BibitemOpen
  \bibfield  {author} {\bibinfo {author} {\bibfnamefont {J.~E.}\ \bibnamefont
  {Avron}}, \bibinfo {author} {\bibfnamefont {M.}~\bibnamefont {Fraas}},
  \bibinfo {author} {\bibfnamefont {G.~M.}\ \bibnamefont {Graf}}, \ and\
  \bibinfo {author} {\bibfnamefont {O.}~\bibnamefont {Kenneth}},\ }\href
  {\doibase 10.1088/1367-2630/13/5/053042} {\bibfield  {journal} {\bibinfo
  {journal} {New J. Phys.}\ }\textbf {\bibinfo {volume} {13}},\ \bibinfo
  {pages} {053042} (\bibinfo {year} {2011})}\BibitemShut {NoStop}%
\bibitem [{\citenamefont {Mahan}(1993)}]{Mahan}%
  \BibitemOpen
  \bibfield  {author} {\bibinfo {author} {\bibfnamefont {G.~D.}\ \bibnamefont
  {Mahan}},\ }\href@noop {} {\emph {\bibinfo {title} {{Many-Particle
  Physics}}}},\ \bibinfo {edition} {2nd}\ ed.\ (\bibinfo  {publisher}
  {Plenum},\ \bibinfo {address} {New York, N.Y.},\ \bibinfo {year}
  {1993})\BibitemShut {NoStop}%
\bibitem [{\citenamefont {Redlich}\ and\ \citenamefont
  {Wijewardhana}(1985)}]{Redlich1985}%
  \BibitemOpen
  \bibfield  {author} {\bibinfo {author} {\bibfnamefont {A.~N.}\ \bibnamefont
  {Redlich}}\ and\ \bibinfo {author} {\bibfnamefont {L.~C.~R.}\ \bibnamefont
  {Wijewardhana}},\ }\href {\doibase 10.1103/PhysRevLett.54.970} {\bibfield
  {journal} {\bibinfo  {journal} {Phys. Rev. Lett.}\ }\textbf {\bibinfo
  {volume} {54}},\ \bibinfo {pages} {970} (\bibinfo {year} {1985})}\BibitemShut
  {NoStop}%
\bibitem [{\citenamefont {Dunne}()}]{dunne1999}%
  \BibitemOpen
  \bibfield  {author} {\bibinfo {author} {\bibfnamefont {G.~V.}\ \bibnamefont
  {Dunne}},\ }\href@noop {} {\ }\Eprint {http://arxiv.org/abs/arXiv:902115
  (1999)} {arXiv:902115 (1999)} \BibitemShut {NoStop}%
\bibitem [{\citenamefont {Coleman}\ and\ \citenamefont
  {Hill}(1985)}]{Coleman1985}%
  \BibitemOpen
  \bibfield  {author} {\bibinfo {author} {\bibfnamefont {S.}~\bibnamefont
  {Coleman}}\ and\ \bibinfo {author} {\bibfnamefont {B.}~\bibnamefont {Hill}},\
  }\href {\doibase 10.1016/0370-2693(85)90883-4} {\bibfield  {journal}
  {\bibinfo  {journal} {Phys. Lett. B}\ }\textbf {\bibinfo {volume} {159}},\
  \bibinfo {pages} {184} (\bibinfo {year} {1985})}\BibitemShut {NoStop}%
\bibitem [{Note22()}]{Note22}%
  \BibitemOpen
  \bibinfo {note} {Violations of this condition \cite {us} play a role similar
  to that of finite temperatures in Hamiltonian settings \cite {dunne1999} and
  may compromise the form of the Chern-Simons theory.}\BibitemShut {Stop}%
\bibitem [{Note23()}]{Note23}%
  \BibitemOpen
  \bibinfo {note} {Despite the lack of thermal symmetry \cite {Sieberer2015},
  the action \protect \textup {\hbox {\mathsurround \z@ \protect \normalfont
  (\ignorespaces \ref {eq:Keld}\unskip \@@italiccorr )}} is invariant under
  another discrete transformation, namely {$\psi _{\pm }^{\protect \tmspace
  +\thinmuskip {.1667em}} \to i\psi ^{\dagger }_{\mp }$}, {$\psi ^{\dagger
  }_{\pm } \to i\psi ^{\protect \tmspace +\thinmuskip {.1667em}}_{\mp }$}, {$S
  \to -S^*$}, where the latter symbol denotes complex conjugation of the
  coefficients of the action. This is the field theoretic counterpart of the
  action of Hermitian conjugation on the Liouvillian. Under this
  transformation, {$S[A_{\pm }] \to -S^*[A_{\mp }]$}, from which the conditions
  on {$M^{IJ}$} follow.}\BibitemShut {Stop}%
\bibitem [{\citenamefont {Altland}\ and\ \citenamefont
  {Simons}(2010)}]{Altland2010Condensed}%
  \BibitemOpen
  \bibfield  {author} {\bibinfo {author} {\bibfnamefont {A.}~\bibnamefont
  {Altland}}\ and\ \bibinfo {author} {\bibfnamefont {B.~D.}\ \bibnamefont
  {Simons}},\ }\href@noop {} {\emph {\bibinfo {title} {{Condensed Matter Field
  Theory}}}},\ \bibinfo {edition} {2nd}\ ed.\ (\bibinfo  {publisher} {Cambridge
  University Press},\ \bibinfo {year} {2010})\BibitemShut {NoStop}%
\bibitem [{Note24()}]{Note24}%
  \BibitemOpen
  \bibinfo {note} {While the quantization of coupling constants in non-abelian
  CS theory is a straightforward consequence of gauge invariance, the situation
  in abelian theories is somewhat more tricky \cite {Tong2016} and depends on
  the topology of the integration manifold.}\BibitemShut {Stop}%
\bibitem [{\citenamefont {Altland}\ and\ \citenamefont
  {Egger}(2009)}]{Altland2009}%
  \BibitemOpen
  \bibfield  {author} {\bibinfo {author} {\bibfnamefont {A.}~\bibnamefont
  {Altland}}\ and\ \bibinfo {author} {\bibfnamefont {R.}~\bibnamefont
  {Egger}},\ }\href@noop {} {\bibfield  {journal} {\bibinfo  {journal} {Phys.
  Rev. Lett.}\ }\textbf {\bibinfo {volume} {102}} (\bibinfo {year}
  {2009})}\BibitemShut {NoStop}%
\bibitem [{\citenamefont {Stone}(1991)}]{Stone1991}%
  \BibitemOpen
  \bibfield  {author} {\bibinfo {author} {\bibfnamefont {M.}~\bibnamefont
  {Stone}},\ }\href {\doibase https://doi.org/10.1016/0003-4916(91)90177-A}
  {\bibfield  {journal} {\bibinfo  {journal} {Ann. Phys.}\ }\textbf {\bibinfo
  {volume} {207}},\ \bibinfo {pages} {38 } (\bibinfo {year}
  {1991})}\BibitemShut {NoStop}%
\bibitem [{\citenamefont {Avron}\ \emph
  {et~al.}(2012{\natexlab{a}})\citenamefont {Avron}, \citenamefont {Fraas},\
  and\ \citenamefont {Graf}}]{Avron2012}%
  \BibitemOpen
  \bibfield  {author} {\bibinfo {author} {\bibfnamefont {J.~E.}\ \bibnamefont
  {Avron}}, \bibinfo {author} {\bibfnamefont {M.}~\bibnamefont {Fraas}}, \ and\
  \bibinfo {author} {\bibfnamefont {G.~M.}\ \bibnamefont {Graf}},\ }\href@noop
  {} {\bibfield  {journal} {\bibinfo  {journal} {J. Stat. Phys.}\ }\textbf
  {\bibinfo {volume} {148}},\ \bibinfo {pages} {800} (\bibinfo {year}
  {2012}{\natexlab{a}})}\BibitemShut {NoStop}%
\bibitem [{\citenamefont {Avron}\ \emph
  {et~al.}(2012{\natexlab{b}})\citenamefont {Avron}, \citenamefont {Fraas},
  \citenamefont {Graf},\ and\ \citenamefont {Grech}}]{Avron2012b}%
  \BibitemOpen
  \bibfield  {author} {\bibinfo {author} {\bibfnamefont {J.}~\bibnamefont
  {Avron}}, \bibinfo {author} {\bibfnamefont {M.}~\bibnamefont {Fraas}},
  \bibinfo {author} {\bibfnamefont {G.~M.}\ \bibnamefont {Graf}}, \ and\
  \bibinfo {author} {\bibfnamefont {P.}~\bibnamefont {Grech}},\ }\href@noop {}
  {\bibfield  {journal} {\bibinfo  {journal} {Commun. Math. Phys.}\ }\textbf
  {\bibinfo {volume} {314}} (\bibinfo {year} {2012}{\natexlab{b}})}\BibitemShut
  {NoStop}%
\bibitem [{\citenamefont {Albert}\ \emph {et~al.}(2016)\citenamefont {Albert},
  \citenamefont {Bradlyn}, \citenamefont {Fraas},\ and\ \citenamefont
  {Jiang}}]{Fraas2016}%
  \BibitemOpen
  \bibfield  {author} {\bibinfo {author} {\bibfnamefont {V.~V.}\ \bibnamefont
  {Albert}}, \bibinfo {author} {\bibfnamefont {B.}~\bibnamefont {Bradlyn}},
  \bibinfo {author} {\bibfnamefont {M.}~\bibnamefont {Fraas}}, \ and\ \bibinfo
  {author} {\bibfnamefont {L.}~\bibnamefont {Jiang}},\ }\href {\doibase
  10.1103/PhysRevX.6.041031} {\bibfield  {journal} {\bibinfo  {journal} {Phys.
  Rev. X}\ }\textbf {\bibinfo {volume} {6}},\ \bibinfo {pages} {041031}
  (\bibinfo {year} {2016})}\BibitemShut {NoStop}%
\bibitem [{\citenamefont {Ma}\ \emph {et~al.}(2019)\citenamefont {Ma},
  \citenamefont {Saxberg}, \citenamefont {Owens}, \citenamefont {Leung},
  \citenamefont {Lu}, \citenamefont {Simon},\ and\ \citenamefont
  {Schuster}}]{B02Schuster2019}%
  \BibitemOpen
  \bibfield  {author} {\bibinfo {author} {\bibfnamefont {R.}~\bibnamefont
  {Ma}}, \bibinfo {author} {\bibfnamefont {B.}~\bibnamefont {Saxberg}},
  \bibinfo {author} {\bibfnamefont {C.}~\bibnamefont {Owens}}, \bibinfo
  {author} {\bibfnamefont {N.}~\bibnamefont {Leung}}, \bibinfo {author}
  {\bibfnamefont {Y.}~\bibnamefont {Lu}}, \bibinfo {author} {\bibfnamefont
  {J.}~\bibnamefont {Simon}}, \ and\ \bibinfo {author} {\bibfnamefont {D.~I.}\
  \bibnamefont {Schuster}},\ }\href@noop {} {\bibfield  {journal} {\bibinfo
  {journal} {Nature}\ }\textbf {\bibinfo {volume} {566}},\ \bibinfo {pages}
  {51} (\bibinfo {year} {2019})}\BibitemShut {NoStop}%
\bibitem [{\citenamefont {Tonielli}\ \emph {et~al.}(tion)\citenamefont
  {Tonielli}, \citenamefont {Budich}, \citenamefont {Altland},\ and\
  \citenamefont {Diehl}}]{us}%
  \BibitemOpen
  \bibfield  {author} {\bibinfo {author} {\bibfnamefont {F.}~\bibnamefont
  {Tonielli}}, \bibinfo {author} {\bibfnamefont {J.~C.}\ \bibnamefont
  {Budich}}, \bibinfo {author} {\bibfnamefont {A.}~\bibnamefont {Altland}}, \
  and\ \bibinfo {author} {\bibfnamefont {S.}~\bibnamefont {Diehl}},\
  }\href@noop {} {\  (\bibinfo {year} {in preparation})}\BibitemShut {NoStop}%
\end{thebibliography}%

\end{document}